\documentclass[aps,pra,superscriptaddress,reprint, twocolumn, amsmath, amssymb, a4, notitlepage,longbibliography]{revtex4-1}
\usepackage{graphicx}
\usepackage{amsmath,amssymb}
\usepackage{dcolumn}
\usepackage{mathrsfs}
\usepackage{physics}
\usepackage{float}
\usepackage{booktabs}
\usepackage{svg}
\usepackage{comment}
\usepackage{hyperref}
\usepackage{color}

\begin{document}
\title{Modeling Quantum Noise in Nanolasers using Markov Chains}
\date{\today}
	
\author{Matias Bundgaard-Nielsen}
\affiliation{Department of Electrical and Photonics Engineering, Technical University of Denmark, Building 343, 2800 Kongens Lyngby, Denmark}
\affiliation{NanoPhoton - Center for Nanophotonics, Technical University of Denmark, Building 343, 2800 Kongens Lyngby, Denmark}
\author{Gian Luca Lippi}
\affiliation{Université Côte d'Azur, CNRS, Institut de Physique de Nice, 17 rue Julien Lauprêtre, 06200
Nice, France}
\author{Jesper M\o rk}
\email[]{jesm@dtu.dk}
\affiliation{Department of Electrical and Photonics Engineering, Technical University of Denmark, Building 343, 2800 Kongens Lyngby, Denmark}
\affiliation{NanoPhoton - Center for Nanophotonics, Technical University of Denmark, Building 343, 2800 Kongens Lyngby, Denmark}

\begin{abstract}
The random nature of spontaneous emission leads to unavoidable fluctuations in a laser's output. This is often included through random Langevin forces in laser rate equations, but this approach falls short for nanolasers. In this paper, we show that the laser quantum noise can be quantitatively computed for a very broad class of lasers by starting from simple and intuitive rate equations and merely assuming that the number of photons and excited electrons only takes discrete values. While the approach has seen previous success, we here derive it rigorously from an open quantum system master equation, whereas it was previously introduced only on phenomenological grounds. We further show that in the many-photon limit, the model simplifies to Langevin equations. We perform an extensive comparison of different approaches for computing quantum noise in lasers, identifying the best approach for different system sizes, ranging from nanolasers to macroscopic lasers, and different levels of excitation, i.e., cavity photon number. In particular, we show that below the laser threshold, stochastic fluctuations in the numerical solution to the Langevin equations can drive populations to unphysical negative values, requiring the introduction of population bounds, which in turn skew the noise statistics, leading to inaccuracies. The Laser Markov Chain model, on the other hand, is accurate for all pump values and laser sizes when collective emitter effects are excluded.

\end{abstract}
\maketitle
 
 \section{Introduction}
Spontaneous emission prevents a laser from ever reaching true equilibrium, acting as a persistent perturbation to the laser state. This leads to inevitable quantum noise in the laser output power and finite laser linewidth \cite{Coldren1997}. In macroscopic lasers, which reach a very high average number of photons in the laser cavity, the domination of stimulated over spontaneous emission can lead to extremely narrow optical linewidths. In nanolasers, the photon population is inherently smaller and the impact of noise is larger \cite{Rice1994PhotonAnalogy,Bjork1991AnalysisEquations,Mork2020SqueezingConfinement}. This is a major challenge in the applications of nanolasers, where low noise operation is critical. These applications include: on-chip communication \cite{Sun2015Single-chipLight}, programmable photonic integrated circuits \cite{Bogaerts2020ProgrammableCircuits}, sensing \cite{Ge2013ExternalBiosensor,Zhang2018ApplicationsCavities,Ma2019ApplicationsNanolasers}, as well as quantum technology \cite{Carolan2015UniversalOptics,Ma2019ApplicationsNanolasers,Wenzel2021SemiconductorRevisited}. The definition of the laser threshold itself is also an issue since \cite{Bjork1991LasersMicrocavities,Jin1994Photon-numberRegime,Bjork1994DefinitionThreshold,Ning2013WhatThreshold, Kreinberg2017EmissionCoupling,Lohof2018DelayedMedia,Takemura2019LasingLasers,Kreinberg2020ThresholdlessFactors,Mork2018RateEmitters,Khurgin2021HowThreshold,Lippi2022PhaseNanolasers,Carroll2021ThermalDevices,Saldutti2024TheNanolasers}, in a nanolaser, a large fraction of the spontaneous emission is funnelled into the lasing mode and the laser operates in a cavity-QED regime rather than in a thermodynamic limit \cite{Rice1994PhotonAnalogy,Takemura2021Low-Analogy}. Research is being conducted to understand and model quantum noise in nanolasers, with the ultimate objective of reducing it. 

Recently, a stochastic approach to simulating the quantum noise in nanolasers has shown promising results \cite{Roy-Choudhury2009QuantumLasers,Roy-Choudhury2010QuantumDiodes,Lebreton2013StochasticallyNanolasers,Puccioni2015StochasticLasing,Mork2018RateEmitters,Andre2020,Bundgaard-Nielsen2023StochasticNanolaser, Takemura2019LasingLasers,Lippi2021AmplifiedNanolasers,Puccioni2024StatisticsNanolasers, Bundgaard-Nielsen2025SimpleNanolasers}. The approach accurately reproduced photon statistics and laser linewidths calculated using full quantum mechanical master equations (MEs) for microscopic nanolasers with a few emitters \cite{Bundgaard-Nielsen2023StochasticNanolaser,Bundgaard-Nielsen2025SimpleNanolasers}. It also agreed with the well-known Langevin approach in the macroscopic limit of lasers containing thousands of emitters \cite{Bundgaard-Nielsen2025SimpleNanolasers}. Importantly, the stochastic approach bridges the mesoscopic regime with ten to hundreds of emitters, where ME calculations become numerically infeasible and Langevin approaches break down due to large noise fluctuations. While alternative methods exist to extend the computational reach beyond direct ME solutions, they all face significant challenges in this regime. Quantum Monte Carlo trajectory methods \cite{Mlmer1993MonteOptics} extend the range of feasibility by evolving a state vector instead of a density matrix, but still suffer exponential scaling. Similarly, the so-called Permutational Invariant Quantum Solver \cite{Shammah2018OpenInvariance} reduces the exponential scaling from $\mathcal{O}(4^N)$ to $\mathcal{O}(N^4)$ where $N$ is the number of emitters, but as we will see, it does not allow simulations to be extended beyond 10 emitters. Tensor network methods and matrix product states \cite{Strathearn2018EfficientOperators,Jrgensen2019ExploitingIntegrals, Finsterholzl2020UsingTime-Evolution,Cygorek2024ACE:Tensors,Link2024OpenContraction} reduce the Hilbert space by truncating the eigenvalues of a decomposed state vector, yet remain highly complex and have not yet been extended to nanolasers. Cumulant expansion methods \cite{Kubo1962GeneralizedCumulantExpansion,Drechsler2022RevisitingNanolasers,Plankensteiner2022quantumcumulantsjl} truncate higher-order quantum correlations in larger systems and are very promising, but require a careful choice of truncation order, and have not been used to compute many-emitter laser linewidths.

Recent advancements in the field of nanofabrication have led to a steady approach to the mesoscopic regime \cite{Dimopoulos2022Electrically-DrivenThreshold}. Consequently, the development of an accurate model in this regime is of paramount importance.

However, while inspired by chemical reaction modeling \cite{gillespie_general_1976,Gillespie1977ExactReactions, Gillespie2000TheEquation,Gillespie2007StochasticKinetics}, no formal derivation connecting the stochastic laser approach to the underlying ME has been established, and its origin has so far only been phenomenological. In this paper we demonstrate that the stochastic description constitutes a Markov chain, where the instantaneous values of the variables encode the entire past history of the evolution, regardless of the individual time scales involved \cite{tolver_markov_chains} and derive it from the ME. We also show that in the many-photon limit, this Markov chain, which we now denote the Laser Markov Chain (LMC), simplifies to the Langevin rate equations (LRE). With the formal connection, we can more easily understand the regimes of validity of the LMC, but also of the LRE, as the derivation of the latter from the former sheds light on the underlying approximations. Furthermore, it provides a simple and intuitive physical picture of the origin of quantum noise, which may aid further insights and new developments.

\begin{figure}[!ht]
    \centering
    \includegraphics[width=1\linewidth]{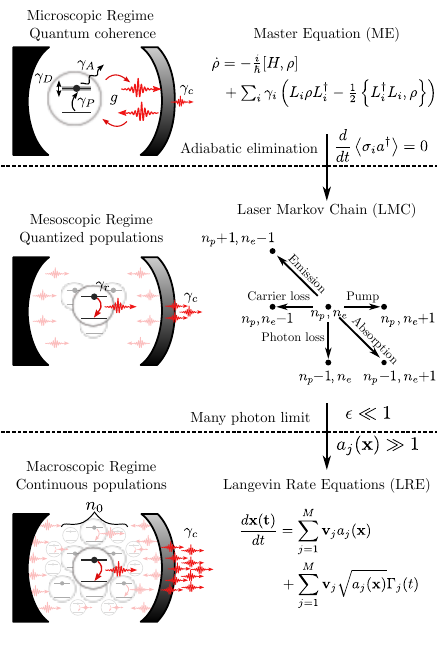}
    \caption{Sketch of the model considered: A cavity containing $n_0$ emitters, which experience various decay and decoherence processes, labeled with $\gamma_i$ (see Table \ref{tab:variables} for the definitions). We show the system in various regimes together with the appropriate model for that regime. MEs capture quantum coherence, but their feasibility for numerical simulation is limited to microscopic cavities. The LMC describes quantized populations well, while LRE are more efficient for large systems where the laser noise is Gaussian (white noise). In the figure $\expval{\sigma_i a^\dagger}$ denotes the average emitter-photon polarization (see Eq.~\eqref{eq:polarization}), $\epsilon$ is the fractional population change due to a stochastic event (Eq.~\eqref{eq:tau_leaping_step}), and $a_j(\mathbf{x})$ represent the rates of the stochastic events (Eq.~\eqref{eq:cme}).}
    \label{fig:cavity_sketch}
\end{figure}

Fig.~\ref{fig:cavity_sketch} shows a sketch of the model considered in this paper, together with an overview of the micro-, meso-, and macroscropic regimes and how the ME, LMC, and LRE are connected through a series of approximations requiring larger and larger systems. In the following section, we introduce the model considered in the paper and derive the well-known laser rate equations. In section III, we introduce Gillespie's method, which gives the necessary context to the LMC, which is introduced in section IV. In section V, we compare the ME, the LMC, and the LRE numerically over a broad range of parameters. In section VI, we study the time vs accuracy of the models to highlight their different regimes of applicability. Finally, in Section VII, we conclude and provide an outlook.

\begin{table}[htbp]
\centering
\caption{Summary of variables and their definitions}
\label{tab:variables}
\begin{tabular}{|c|c|}
\toprule
\hline
\textbf{Variable} & \textbf{Description} \\\hline
\midrule
$n_p$ & Photon population ($n_p  = \langle a^\dagger a \rangle$) \\ \hline
$n_e$ & Excited emitter population ($n_e = \sum_i \langle \sigma_i^\dagger \sigma_i\rangle$) \\ \hline
$n_0$ & Number of two-level emitters in the cavity \\\hline

$g$, $g_i$ & Light-matter coupling strength \\\hline
\midrule
$\gamma_c$ & Cavity decay rate \\\hline
$\gamma_A$ & Background decay rate per emitter \\\hline
$\gamma_P$ & Pump rate per emitter \\\hline
$\gamma_D$ & Pure dephasing rate per emitter \\\hline
$\gamma_r$ & Radiative decay rate ($\gamma_r = 4g^2/\gamma_\perp$) \\\hline
$\gamma_\perp$ & Total dephasing rate ($\gamma_\perp = \gamma_P+\gamma_A+\gamma_D+\gamma_c$) \\ \hline
\midrule
\end{tabular}
\end{table}

\section{System and Model \label{sec:system_and_model}}


To model our laser, we consider $n_0$ two-level systems all coupled to a single optical cavity. This is well described by the ME \cite{Mu1992One-atomLasers,Loffler1997SpectralLaser,Moelbjerg2013DynamicalEmitters} for the density operator $\rho$:
\begin{equation}
\begin{aligned}
    \frac{\partial \rho}{\partial t} &=-i[H, \rho]+\gamma_c \mathcal{D}_a(\rho)+\sum_i \gamma_A \mathcal{D}_{\sigma_i}(\rho)
\\ &+\sum_i \gamma_P \mathcal{D}_{{\sigma_i}^\dagger}(\rho) +\sum_i \gamma_D \mathcal{D}_{{\sigma_i}^\dagger \sigma}(\rho). \label{eq:me}
\end{aligned}
\end{equation}
Here, $H$ is the typical Jaynes-Cummings Hamiltonian for the $n_0$ emitters, the photons in the cavity, and their interaction: $H= \hbar \omega_c a^\dagger a + \sum_i \omega_{e_i} \sigma_i^\dagger \sigma_i +   \sum_i g_i (a^\dagger \sigma_i + a \sigma_i^\dagger)$, where $a$ is the annihilation operator of the field in the cavity mode and $\sigma_i = \ket{g}_i\bra{e}_i$, is the lowering operator of the i-th emitter. We assume all emitters to be identical and in resonance with the cavity, thus the light-matter coupling strength $g_i=g$ is identical for all emitters and $\omega_c =\omega_e = \omega_{e_i}$. In the ME, we account for dissipation through Lindblad terms with  $\mathcal{D}_A(\rho)=A \rho A^{+}-\frac{1}{2}\left(A^{\dagger} A \rho+\rho A^{\dagger} A\right)$, where  $\gamma_c$ describes the cavity decay rate, $\gamma_A$ the background decay per emitter, $\gamma_P$ the pump rate per emitter modeled as an inverse decay term / inhomogeneous pumping \cite{Mu1992One-atomLasers, Loffler1997SpectralLaser,Benson1999Master-equationLaser,Ritter2010EmissionLaser}, and $\gamma_D$ the pure dephasing rate per emitter, arising from, e.g., electron-electron and electron-phonon scattering \cite{Strauf2011SingleNanolaser}. 

Before deriving the Laser Markov Chain model from the ME, we derive the laser rate equations to highlight the essential differences. The typical approach is to derive the equations of motion for the operator expectation values $n_p = \expval{a^\dagger a}$ and $n_e = \sum_i \expval{\sigma_i^\dagger \sigma_i}$, describing the photon and emitter populations, respectively. This way we obtain ( see e.g., \cite{Lorke2013TheoryTheory,Moelbjerg2013DynamicalEmitters}):
\begin{align}
&\frac{d}{d t}\left\langle a^{\dagger} a\right\rangle=\sum_{i=1}^N 2 g \operatorname{Im}\left[\left\langle\sigma_i a^{\dagger}\right\rangle\right]-\gamma_c\left\langle a^{\dagger} a\right\rangle, \\
&\frac{d}{d t}\left\langle\sigma_i^{\dagger} \sigma_i\right\rangle=\gamma_P\left\langle\sigma_i \sigma_i^{\dagger}\right\rangle-2 g \operatorname{Im}\left[\left\langle\sigma_i a^{\dagger}\right\rangle\right]-\gamma_A \left\langle\sigma_i^{\dagger} \sigma_i\right\rangle .
\end{align}
Both equations couple to the off-diagonal polarization $\left\langle\sigma_i a^{\dagger}\right\rangle$, which follows the equation of motion:
\begin{equation}
\begin{aligned}
\frac{d}{d t}\left\langle\sigma_i a^{\dagger}\right\rangle &=i g\left(\left\langle\sigma_i^{\dagger} \sigma_i a a^{\dagger}\right\rangle-\left\langle\sigma_i \sigma_i^{\dagger} a^{\dagger} a\right\rangle\right) \\ &- \frac{1}{2} \gamma_\perp \left\langle\sigma_i a^{\dagger}\right\rangle \label{eq:polarization}
\end{aligned}
\end{equation}
where we defined the total dephasing rate $\gamma_\perp = \gamma_P+\gamma_A+\gamma_D+\gamma_c$. The problem now is that this equation involves higher-order correlations, such as $\left\langle\sigma_i^{\dagger} \sigma_i a a^{\dagger}\right\rangle$. To close the otherwise infinite set of equations of motion, we make two approximations. First, the cumulant expansion, where we approximate \cite{Fricke1996TransportFormalism,Gartner2011Two-levelTransition,Gies2007SemiconductorLasers,Lorke2013TheoryTheory,Moelbjerg2013DynamicalEmitters} $\left\langle\sigma_i^{\dagger} \sigma_i a a^{\dagger}\right\rangle=\left\langle\sigma_i^{\dagger} \sigma_i\right\rangle\left\langle a a^{\dagger}\right\rangle$ and $\left\langle\sigma_i \sigma_i^{\dagger} a^{\dagger} a\right\rangle=$ $\left\langle\sigma_i \sigma_i^{\dagger}\right\rangle\left\langle a^{\dagger} a\right\rangle$. This approximation amounts to ignoring the correlations between the emitter and photon populations; well-justified if $4g < \gamma_\perp$. When this approximation holds, we can also take the adiabatic elimination and assume that $d/dt \left\langle\sigma_i a^{\dagger}\right\rangle = 0$: the emitter polarization follows the photon and emitter populations. This approximation implies that Rabi oscillations and collective effects such as super- and sub-radiant emission processes \cite{Maki1989InfluenceSuperfluorescence,Drummond1991QuantumMedium,Auffeves2011FewIndividualization,Gies2015PhotonCavity,Jahnke2016GiantNanolasers} cannot be described by the model. With these approximations, we can insert the expression for $\left\langle\sigma_i a^{\dagger}\right\rangle$ in Eq.~\eqref{eq:polarization} and rework the rate equations to obtain:
\begin{align}
      &\frac{d n_p}{dt} = \gamma_r (2n_e-n_0)n_p + \gamma_r n_e- \gamma_c n_p, \label{eq:na} \\
      &\frac{d n_e}{dt} = \gamma_P (n_0-n_e) - \gamma_r (2n_e-n_0)n_p -\gamma_r n_e - \gamma_A n_e, \label{eq:ne}
\end{align}
which are the well-known laser rate equations, see e.g. \cite{Mork2018RateEmitters,Yokoyama1989RateLasers}. Here, $\gamma_r = 4g^2/\gamma_\perp$ is the radiative decay rate. We refer to Table \ref{tab:variables} for a definition of all variables. 

We now move on to derive the stochastic LMC from the ME and show that this probabilistic view similarly simplifies to the above rate equation in the many-photon limit. This also means that the LMC can be inferred from the rate equations. This is how the Markov chains have been employed previously \cite{Puccioni2015StochasticLasing,Mork2018RateEmitters,Andre2020,Bundgaard-Nielsen2023StochasticNanolaser, Takemura2019LasingLasers,Lippi2021AmplifiedNanolasers,Puccioni2024StatisticsNanolasers, Bundgaard-Nielsen2025SimpleNanolasers}, and in Sec.~\ref {sec:markov_chain_from_re} we give a concrete example on how to infer the stochastic processes in the Markov chain from eqs.~\eqref{eq:na}-\eqref{eq:ne}. This gives a simple recipe for translating rate equations to Laser Markov Chains, which do not have the problems of the Langevin approach highlighted in Sec .~\ref{sec:numerical}.   

\section{Gillespie's method \label{sec:GFRM}}
As already mentioned above, modeling discrete populations is well understood and explored in chemistry \cite{Gillespie2007StochasticKinetics}. Large chemical systems with many billions of reactants are well-described by deterministic rate equations. However, the effects of having a discrete number of reactants in a chemical system are particularly important when the total number of reactants is low, as in small biochemical systems, where stochastic fluctuations occur  \cite{gillespie_general_1976,Gillespie1977ExactReactions,Gillespie1992AEquation,Gillespie2007StochasticKinetics}. As we will see, this is similar to how the classical laser rate equations provide a good description of large semiconductor lasers, while in nanolasers one needs to take into account the discrete nature of photons and emitters. 

Before deriving the LMC model from the ME, we briefly introduce the relevant theory from chemical modeling, notably Gillespie's method \cite{Gillespie2007StochasticKinetics}.
In the most general terms, we are considering $N$ distinct populations $\mathbf{x}(t) \equiv\left(x_1(t), \ldots, x_N(t)\right)$, which interact through $M$ processes $\{R_1,...,R_M\}$. Each process $R_j$ is associated with a population change $\mathbf{v}_j = \left(v_{1j}, \ldots, v_{Nj}\right)$, where $v_{ij}$ is the change in the $x_i$  population caused by the $R_j$ process. Thus, if the system is in state $\mathbf{x}$ and
the $R_j$ process occurs, the system jumps to state $\mathbf{x} + \mathbf{v}_j$. The rate of a given process is given by the propensity function $a_j(\mathbf{x})$, where $a_j(\mathbf{x}) dt$ is the probability of the process $R_j$ occurring in the time interval $dt$ given the initial state $\mathbf{x}$. With these assumptions, the probability of being in the state $\mathbf{x}$ at time $t$, denoted by $P(\mathbf{x},t)$, can be rigorously derived from probability theory to be governed by \cite{gillespie_general_1976}:
\begin{equation}
    \frac{\partial P\left(\mathbf{x}, t\right)}{\partial t}=\sum_{j=1}^M\left[a_j\left(\mathbf{x}-\boldsymbol{v}_j\right) P\left(\mathbf{x}-\boldsymbol{v}_j, t \right)-a_j(\mathbf{x}) P\left(\mathbf{x}, t\right)\right] \label{eq:cme}
\end{equation} 
Notice that the first term is a "gain-like" term that describes the probability flow into the state $P(\mathbf{x},t)$ from neighboring states $P(\mathbf{x}-\mathbf{v}_j,t)$: it holds contributions from all processes leading to that particular state. Similarly, the second term is a "loss-like" term that describes the total probability flow away from the state $P(\mathbf{x},t)$ into states $P(\mathbf{x}+\mathbf{v}_j,t)$. 

Much like the ME, Eq.~\eqref{eq:cme} quickly grows intractable when the size of the system increases, due to the sheer number of possible states the system can be in. To combat this, Gillespie derived and developed a Monte-Carlo simulation scheme that samples the state space $P(\mathbf{x},t)$ appropriately at a much reduced numerical cost \cite{gillespie_general_1976}. The simulation scheme that generates trajectories $\mathbf{x}_i(t)$ which evolve according to the probability distribution $P(\mathbf{x},t)$, relies on another probabilistic quantity $P(\tau,j|\mathbf{x},t)$, which describes the probability that the process (transition) $R_j$ will occur after a time $\tau$ and is of type $R_j$. Again, using probability theory, it can be shown that $P(\tau,j|\mathbf{x},t)$ is given as \cite{Gillespie1992AEquation}:
\begin{equation}
	P(\tau,j|\mathbf{x},t) = a_j(\mathbf{x}) \exp \left ( \sum_{i=1}^M a_i(\mathbf{x}) \tau \right). 
\end{equation}
Thus, the time $\tau$, at which the process $R_j$ occurs, is exponentially distributed. To simulate the evolution of the trajectory, we thus draw a random $\tau$ and $j$ from this exponential distribution (we will discuss shortly how) and then update the population according to the associated change $\mathbf{v}_j$. 

Gillespie's method can be summarized as the following numerical algorithm:

\begin{enumerate}
\item Initialize the simulation time $t=0$ and the integer-valued population vector
      $\mathbf{x}(0)=\mathbf{x}_0$.

\item For the current state $\mathbf{x}(t)$, evaluate all reaction propensities
      $a_j(\mathbf{x})$ for $j=1,\dots,J$.

\item For each process $j$, draw a random waiting time
\begin{equation}
	\tau_j = \frac{1}{a_j(\mathbf{x})}\ln\!\left(\frac{1}{r_j}\right),
\end{equation}
where $r_j\in(0,1)$ is uniformly distributed.

\item Identify the process $j^\ast$ associated with the smallest waiting time
      $\tau = \min \{ \tau_j \}$.

\item Update the simulation time and populations according to the discrete jump
\begin{align}
	t &\rightarrow t + \tau, \\
	\mathbf{x}(t+\tau) &= \mathbf{x}(t) + \mathbf{v}_{j^\ast},
\end{align}
where $\mathbf{v}_{j^\ast}$ is the population update vector for
process $j^\ast$.

\item Record the updated state $\mathbf{x}(t)$ and repeat from step~2 until the chosen simulation time is reached.
\end{enumerate}


Gillespie's method is thus a Markov chain that explores the state space $P(\mathbf{x},t)$, which follows an equation of the type in Eq.~\eqref{eq:cme}. In the following section, we show that under certain approximations, the ME in Eq.~\eqref{eq:me} takes the form of Eq.~\eqref{eq:cme} and thus can be sampled using Gillespie's method. In Sec.~\ref{sec:markov_chain_from_re}, we also highlight how one can identify the Markov chain in a simple way directly from a rate equation.

\section{Laser Markov Chain \label{sec:LMC}}
In the previous section, we introduced Gillespie's method, which is used in chemical modeling to simulate the evolution of integer-valued populations. In this section, we show that starting from the ME in Eq.~\eqref{eq:me}, we can derive a similar stochastic model. We also show that in the many-photon limit, this model simplifies to the LRE.

\subsection{Derivation}
To derive the LMC from the ME in Eq.~\eqref{eq:me}, we need to reformulate the latter into a form similar to Eq.~\eqref{eq:cme}. The probabilistic evolution in Eq.~\eqref{eq:cme}, describes only diagonal elements of the density matrix. This matches the Scully-Lamb birth-death ME \cite{Scully1967QuantumTheory} of quantum laser theory where the dynamics of the entire emitter population has been eliminated. 
In the following, we derive a Scully-Lamb-like ME which retain the emitter population and highlight the relation between the LMC and ME, explaining the previous success of the LMC \cite{Puccioni2015StochasticLasing,Mork2020SqueezingConfinement,Andre2020,Bundgaard-Nielsen2023StochasticNanolaser,Bundgaard-Nielsen2025SimpleNanolasers}.

We wish to obtain the equations of motion for $\rho_{n_p,n_e}$ which describe the probability of $n_p$ photons and $n_e$ emitters. First, we derive the equation for the matrix element $\rho_{n_p,g;n_p,g} = \langle n_p,g| \rho |n_p, g\rangle$, describing a single emitter in the ground state. By applying the projector $\langle n_p,g| \rho |n_p, g\rangle$ to Eq.~\eqref{eq:me}, we get:
\begin{equation}
\begin{aligned}
    \frac{\partial \rho_{n_p,g;n_p,g}}{\partial t} &=i \sqrt{n_p} g (\rho_{n_p,g;n_p-1,e} - \rho_{n_p-1,e;n_p,g}) \\
&+ \gamma_c (n_p+1) \rho_{n_p+1,g;n_p+1,g}- \gamma_c n \rho_{n_p,g;n_p,g} \\
&+ \gamma_A \rho_{n_p,e;n_p,e}-\gamma_P \rho_{n_p,g;n_p,g}  \label{eq:rho_ng_nonsimple}
\end{aligned}
\end{equation}
For matrix elements $\rho_{n_p,e;n_p,e}$, one similarly arrives at:
\begin{equation}
\begin{aligned}
    \frac{\partial \rho_{n_p,e;n_p,e}}{\partial t} &=-i \sqrt{n_p+1} g ( \rho_{n_p+1,g;n_p,e} -\rho_{n_p,e;n_p+1,g})  \\
 &+ \gamma_c (n_p+1) \rho_{n_p+1,e;n_p+1,e}- \gamma_c n \rho_{n_p,e;n_p,e} \\
&- \gamma_A \rho_{n_p,e;n_p,e} + \gamma_P \rho_{n_p,g;n_p,g} \label{eq:rho_ne_nonsimple}
\end{aligned}
\end{equation}

To continue, we need to calculate the off-diagonal elements $\rho_{n_p,g;n_p-1,e}$ and $\rho_{n_p-1,e;n_p,g} = \rho_{n_p,g;n_p-1,e}^\dagger$.  We get:
\begin{equation}
\begin{aligned}
    &\frac{\partial \rho_{n_p,g;n_p-1,e}}{\partial t} = i \sqrt{n_p} g (\rho_{n_p,g;n_p,g} - \rho_{n_p-1,e;n_p-1,e}) \\
&+ \gamma_c \left ( \sqrt{n_p+1}\sqrt{n_p} \rho_{n_p+1,g;n_p,e}  - \frac{(2 n_p-1)}{2} \rho_{n_p,g;n_p-1,e} \right) \\ 
& - \frac{\gamma_A +\gamma_P+\gamma_D}{2} \rho_{n_p,g;n_p-1,e}. \label{eq:offdiagonal}
\end{aligned}
\end{equation}

The off-diagonal elements here play the same role as the polarization $\expval{\sigma a^\dagger}$ plays in the derivation of the laser rate equations in Eq.~\eqref{eq:polarization}. There (section \ref{sec:system_and_model}), the off-diagonal elements are eliminated using a cumulant expansion and the adiabatic elimination $d/dt \left\langle\sigma_i a^{\dagger}\right\rangle = 0$.

In this derivation, the adiabatic elimination amounts to setting $\partial \rho_{n_p,g;n_p-1,e} / \partial t = 0$, to obtain the analytical expression for the equilibrium form of the off-diagonal element $\rho_{n_p,g;n_p-1,e}$; however, we notice the coupling to $\rho_{n_p+1,g;n_p,e}$, in a way similar to how $\expval{\sigma a^\dagger}$ couples to higher order terms. Unlike the cumulant expansion used in section \ref{sec:system_and_model}, here there is no clear interpretation of the role or form of the cumulant expansion. However, assuming: 
\begin{equation}
    \rho_{n_p+1,g;n_p,e} \approx   \frac{(n_p-1) }{\sqrt{n_p^2+n_p}}  \rho_{n_p,g;n_p-1,e} ,\label{eq:matrix_element_approximation}
\end{equation}
we obtain a simplified form for the off-diagonal elements:
\begin{equation}
    \rho_{n_p,g;n_p-1,e} = \frac{2 i \sqrt{n_p} g (\rho_{n_p,g;n_p,g} - \rho_{n_p-1,e;n_p-1,e})}{\gamma_A +\gamma_P+\gamma_D+\gamma_c};
\end{equation}
a comment on the validity of the approximation is offered in 
appendix \ref{sec:validity}. Finally, inserting this expression into Eqs.~\eqref{eq:rho_ng_nonsimple} and \eqref{eq:rho_ne_nonsimple} we get:
\begin{equation}
\begin{aligned}
    \frac{\partial \rho_{n_p,g;n_p,g}}{\partial t} &= \gamma_r n_p (\rho_{n_p-1,e;n_p-1,e} - \rho_{n_p,g;n_p,g}) \\ 
&+ \gamma_c (n_p+1) \rho_{n_p+1,g;n_p+1,g}- \gamma_c n_p \rho_{n_p,g;n_p,g} \\
&+ \gamma_A \rho_{n_p,e;n_p,e}-\gamma_P \rho_{n_p,g;n_p,g} .\label{eq:rhong}
\end{aligned}
\end{equation}
\begin{equation}
\begin{aligned}
    \frac{\partial \rho_{n_p,e;n_p,e}}{\partial t} &= (n_p+1) \gamma_r (\rho_{n_p+1,g;n_p+1,g} -\rho_{n_p,e;n_p,e}) \\ 
&+ \gamma_c (n_p+1) \rho_{n_p+1,e;n_p+1,e}- \gamma_c n_p \rho_{n_p,e;n_p,e} \\
&- \gamma_A \rho_{n_p,e;n_p,e} + \gamma_P \rho_{n_p,g;n_p,g} .\label{eq:rhone}
\end{aligned}
\end{equation}
where again $\gamma_r = 4 g^2 / \gamma_\perp$ (see Table \ref{tab:variables}). We recognize that this corresponds to the radiative decay rate of the nanolaser rate equations \cite{Mork2018RateEmitters,Bundgaard-Nielsen2023StochasticNanolaser}. This suggests that the assumption in, Eq.~\eqref{eq:matrix_element_approximation}, is in fact equivalent to the cumulant expansion. Compared to the cumulant expansion, however, we now have the possibility of a direct comparison and verification of the approximation (see appendix \ref{sec:validity}).

By extending the previous analysis to each of the $n_0$ emitters, we arrive at equations similar to Eqs.~\eqref{eq:rhong} and \eqref{eq:rhone} written for each individual emitter. With the introduction of $\rho_{n_p,n_e}$ describing the sum of all density matrix elements with $n_p$ number of photons and $n_e$ excited emitters and through the use of combinatorics accounting for all the different microscopic emitter configurations (see appendix \ref{app:extending} for details), we arrive at:
\begin{equation}
    \begin{aligned}
        &\frac{\partial  \rho_{n_p,n_e}}{\partial t} = (n_e+1) \gamma_r n_p \rho_{n_p-1,n_e+1} - n_e (n_p+1) \gamma_r \rho_{n_p,n_e} \\
    &+ (n_0-n_e + 1) (n_p+1) \gamma_r \rho_{n_p+1,n_e-1} - \gamma_r n_p (n_0-n_e) \rho_{n_p,n_e} \\  
    &+ \gamma_c (n_p+1)\rho_{n_p+1,n_e} - \gamma_c n_p \rho_{n_p,n_e} \\ 
    &+  (n_e+1) \gamma_A \rho_{n_p,n_e+1} - \gamma_A n_e \rho_{n_p,n_e} \\ 
    &+ \gamma_P (n_0-n_e+1) \rho_{n_p,n_e-1} - (n_0-n_e) \gamma_P \rho_{n_p,n_e} \label{eq:slme}
    \end{aligned}
\end{equation}
The above equation follows exactly the structure of Eq.~\eqref{eq:cme}. In the following section, we show the processes involved in the LMC and thus how the nanolaser can be simulated using Gillespie's method.

We note that while this derivation treats the carriers as simple two-level systems with Markovian dynamics, models involving multiple Markov chains have been formulated to allow for more complex carrier dynamics involving multiple levels \cite{Vallet2019MarkovNanolasers}. In general, such an extension will, however, not be necessary as long as the laser rate equations are a good description of the laser dynamics.

\begin{table*}[!ht]
\centering
\caption{Rates of event types in the laser rate equations. \label{tab:events}}
\begin{tabular}{|c|c|c|}
\hline
\textbf{Event type} & \textbf{Rate, $a_j$} & \textbf{Population change}, $\mathbf{v}_j$ ($n_p$,$n_e$)  \\ \hline
Stimulated emission & $a_{st} = \gamma_r n_e(t) n_p(t)$ & (+1,-1) \\ \hline
Spontaneous emission & $a_{sp} =\gamma_r n_e(t)$ & (+1,-1 ) \\ \hline
Absorption &$a_{ab} = \gamma_r (n_0 - n_e(t)) n_p(t)$ & (-1,+1) \\ \hline
Cavity decay & $a_{c} =\gamma_c n_p(t)$ & (-1,0) \\ \hline
Background decay &  $a_{bg} =\gamma_{A} n_e(t)$ & (0,-1) \\ \hline
Pumping & $a_{p} = \gamma_P (n_0 - n_e(t))$ & (0,+1) \\ \hline
\end{tabular}
\end{table*}

\subsection{Gillespie's method applied to the Laser Markov Chain}
In the previous section, we derived a Scully-Lamb-like ME in Eq.~\eqref{eq:slme}, which governs the evolution of the probability distribution of $\rho_{n_p,n_e}$. Since Eq.~\eqref{eq:slme} follows the exact structure of Eq.~\eqref{eq:cme}, we can efficiently simulate it using Gillespie's method. We can further illustrate this by defining $\textbf{x} = (n_p,n_e)$ and inserting the events from Table \ref{tab:events} into Eq.~\eqref{eq:cme}, see appendix \ref{app:extending}. We find that we recover Eq.~\eqref{eq:slme} line by line. Following Gillespie's method outlined in Section \ref{sec:GFRM} with the events and rates of Table \ref{tab:events}, we can thus efficiently sample the probability distribution outlined in Eq.~\eqref{eq:slme}. For a numerical implementation of the scheme, see the supplementary material of Refs.~\cite{Bundgaard-Nielsen2023StochasticNanolaser,Bundgaard-Nielsen2025SimpleNanolasers} or Ref.\cite{github}. 

This derivation thus shows a formal connection between the ME in Eq.~\eqref{eq:me} and the LMC. This explains the success of the stochastic approach employed in refs.~\cite{Bundgaard-Nielsen2023StochasticNanolaser,Bundgaard-Nielsen2025SimpleNanolasers}, which was numerically observed to compare favorably to the ME in contrast with a typical Langevin approach.

In previous works \cite{Puccioni2015StochasticLasing,Mork2018RateEmitters,Andre2020,Bundgaard-Nielsen2023StochasticNanolaser, Takemura2019LasingLasers,Lippi2021AmplifiedNanolasers,Puccioni2024StatisticsNanolasers, Bundgaard-Nielsen2025SimpleNanolasers}, this formal connection had not been established; instead, the events and rates have been inferred from an already derived rate equation. In the following sections, we show how one obtains the rate equations from the LMC in the many-photon limit, thus connecting all the methods for including noise in nanolasers discussed in this paper. We also show that inferring the events and rates from a rate equation is equivalent to the above derivation, which makes the adoption of the LMC methods very simple.  

\subsection{Intermediate Photon Limit  - Tau-Leaping \label{sec:tau_leaping}}
If the mean number of photons is large, it can be computationally inefficient to follow all the individual population changes, since several of the rates (see Table \ref{tab:events}) scale with the photon number, leading to extremely small time steps. The number of samples required to achieve accurate photon statistics, such as the $g^{(2)}(0)$-function, can thus get very large, since one needs to simulate over time durations that are long compared to the characteristic time scales of the laser, such as the carrier lifetime. Instead, if one assumes that during a time window $\tau$, the rates $a_j(\mathbf{x})$ are essentially constant, then the number of times each process  $j$ happens is a Poisson random variable with mean $a_j(\mathbf{x}) \tau$ \cite{Gillespie2007StochasticKinetics}.

In other words, if this leap condition is valid, we can approximate the time evolution of the system as \cite{gillespie_general_1976,Gillespie1977ExactReactions,Gillespie2007StochasticKinetics}:
\begin{equation}
    \mathbf{x}(t+\tau) = \mathbf{x}(t) + \sum_j^m\mathcal{P}(a_j(\mathbf{x(t)}) \tau) \mathbf{v}_j \label{eq:tau_leaping}
\end{equation}
where $\mathcal{P}(a_j(\mathbf{x(t)})$ is a random variable drawn from a Poisson distribution with mean and variance $a_j(\mathbf{x(t)})$. From Eq.~\ref{eq:tau_leaping} it is clear that if the changes induced by $\mathcal{P}(a_j(\mathbf{x(t)}) \tau)$ significantly alter the rates $a_j(\mathbf{x(t)})$, then during the step we could no longer assume the rates $\mathcal{P}(a_j(\mathbf{x(t)}) \tau)$ to be constant. As a consequence, the time increment $\tau$ should be chosen accordingly to ensure this does not happen. Note, however, that if $\tau$ is chosen too small, the algorithm quickly becomes numerically inefficient as many of the Poisson draws will result in zero population changes. 

To choose $\tau$ efficiently, we can define another numerical parameter $\epsilon$ and require that the expected fractional population change is bounded such that $| \Delta \mathbf{x} / \mathbf{x}| = |(\mathbf{x}(t+\tau) - \mathbf{x}(t))/\mathbf{x}(t)|<\epsilon$ at each timestep. Ensuring that both the standard deviation and mean number change due to Eq.~\eqref{eq:tau_leaping} is bounded by $\epsilon \mathbf{x}$ translates into \cite{Gillespie2007StochasticKinetics}:
\begin{equation}
    \tau < \min \left( \frac{\epsilon x_i}{v_{i j} a_j} ,\frac{\epsilon^2 x_i^2}{v_{i j}^2 a_j}\right ), \ \ \mathrm{for \ all}\, i\ \mathrm{and} \ j.  \label{eq:tau_leaping_step}
\end{equation}
Generally, we find that $\epsilon \approx$ $0.005$--$0.01$ to achieve accurate convergence in our laser simulations. See appendix \ref{app:numerical_details} for more details on the numerics.  

\subsection{Many Photon Limit - Langevin Equations \label{sec:langevin}}
If we can find a $\tau$ for which the fractional change $\epsilon$ is small \textit{and} $a_j(\mathbf{x})\tau>>1$, meaning that at each leap many events occur, but they do not change the population significantly, we can approximate the Poisson random variables as Gaussian variables. Using the fact that a gaussian variable with mean $m$ and standard deviation $\sigma$ can be written as $\mathcal{N}(m,\sigma^2) = m + \sigma \mathcal{N}(0,1)$, we obtain the time evolution:
\begin{equation}
\mathbf{x}(t+\tau) = \mathbf{x}(t)+\sum_{j=1}^M \mathbf{v}_j a_j(\mathbf{x}) \tau+\sum_{j=1}^M \mathbf{v}_j \sqrt{a_j(\mathbf{x})} \mathcal{N}_j(0,1) \sqrt{\tau} \label{eq:langevin_limit}
\end{equation}
Taking the limit of $\tau \rightarrow 0$ and rearranging, Eq.~\eqref{eq:langevin_limit} becomes:
\begin{equation}
    \frac{d \mathbf{x(t)}}{dt} = \sum_{j=1}^M \mathbf{v}_j a_j(\mathbf{x}) + \sum_{j=1}^M \mathbf{v}_j \sqrt{a_j(\mathbf{x})} \Gamma_j(t) \label{eq:sde}
\end{equation}
where $\Gamma_j(t)$ is an uncorrelated Gaussian white noise which satisfies: $\expval{\Gamma_i(t)\Gamma_j(s)}=\delta_{ij}\delta(t-s)$. Eq.~\ref{eq:sde} has the exact form of a rate equation with stochastic Langevin forces on the right-hand side. Using the events and rates of Table \ref{tab:events}, one recovers Eqs~\eqref{eq:na}-\eqref{eq:ne}, but for the added stochastic force $F_i = \sum_j \mathbf{v}_{j}[i] \sqrt{a_j(\mathbf{x})}\Gamma_j(t)$, where $\mathbf{v}_{j}[i]$ is the i-th index of the j-th population change vector $\mathbf{v}_j$. With this, we can furthermore recover the Langevin diffusion constants defined via: 
\begin{equation}
    \expval{F_i(t)F_j(s)} = \sum_k \mathbf{v}_{k}[i]\mathbf{v}_{k}[j]a_k(\mathbf{x}) \delta(t-s) = 2 D_{ij} \delta(t-s) \label{eq:f_whitenoise}
\end{equation}
i.e., $2 D_{ij} =  \sum_k \mathbf{v}_{k}[i]\mathbf{v}_{k}[j]a_k(\mathbf{x})$. Inserting the values from Table \ref{tab:events} we get: 
\begin{align}
D_{pp} &= \frac12\bigl[\gamma_r n_0 n_p + \gamma_r n_e + \gamma_c n_p\bigr] \label{eq:d_aa},\\
D_{pe} &= -\frac12\,\gamma_r\bigl(n_e + n_0 n_p\bigr)  \label{eq:d_ae},\\
D_{ee} &= \frac12\Bigl[\gamma_r n_0 n_p + \gamma_r n_e + \gamma_P\,(n_0-n_e) + \gamma_A n_e\Bigr], \label{eq:d_ee}
\end{align}
where the subscripts $p$ and $e$ refer to photon and emitter, respectively. The diffusion constants agree with previous derivations \cite{Mork2020SqueezingConfinement}, which were obtained through the usual Langevin approach \cite{Coldren1997}. This derivation shows that the Langevin approach is the many-photon limit of the LMC. Previous results have shown that the LMC is valid in the few-photon limit (when compared with a ME), while the Langevin approach has shown significant deviations \cite{Bundgaard-Nielsen2023StochasticNanolaser,Bundgaard-Nielsen2025SimpleNanolasers}, in line with the present derivation.

This derivation also gives a clear recipe for deriving a LMC from a rate equation, a point on which we expand in the following section.

\subsection{Laser Markov Chain from Rate Equations \label{sec:markov_chain_from_re}}

In sections \ref{sec:LMC}-\ref{sec:langevin}, we have shown how to transition from a ME to the LMC, which can be sampled either via Gillespie's method or the Tau-Leaping approach. We also showed that in the many-photon limit, the LMC simplifies to the Langevin approach.

While our derivation provides theoretical foundation for the LMC, previous works \cite{Roy-Choudhury2009QuantumLasers,Roy-Choudhury2010QuantumDiodes,Puccioni2015StochasticLasing,Mork2018RateEmitters,Andre2020,Bundgaard-Nielsen2023StochasticNanolaser, Takemura2019LasingLasers,Lippi2021AmplifiedNanolasers,Puccioni2024StatisticsNanolasers, Bundgaard-Nielsen2025SimpleNanolasers} developed it phenomenologically by interpreting rate equations as stochastic processes. As shown in the following sections, our current derivation validates the intuitive approach, while providing a solid interpretation of the more accurate modeling of laser quantum noise in the presence of small populations.

The shortcomings of the direct derivation of the LMC from the ME is the lengthy and cumbersome treatment of off-diagonal density matrix elements and combinatorial factors for multiple emitters. The welcome surprise, Eq.~\eqref{eq:sde}, is the possibility of obtaining the events, rates, and associated changes directly from the Langevin rate equation.

In the following, we analyze Eqs.~\eqref{eq:na}-\eqref{eq:ne} with the purpose of reconstructing the events and rates in Table \ref{tab:events}. Each term in the classical rate equations corresponds to a specific microscopic process that changes the photon and emitter populations by discrete amounts. The key insight is that any term of the form $\alpha \cdot f(n_p, n_e)$ in the rate equation represents a process occurring at rate $ f(n_p, n_e)$, where the sign of $\alpha$ indicates whether the population increases or decreases.

For example, if we consider the rate $\gamma_r n_e$, which occurs as the second element in Eq.~\eqref{eq:na} and the third element in Eq.~\eqref{eq:ne} with opposite sign, we can immediately identify $\textbf{v} = (+1,-1)$ with the rate $\gamma_r n_e$. This process is related to spontaneous emission, as shown in Table \ref{tab:events}. 

The rate $\gamma_r (2n_e-n_0)n_p$, occurring as first and second term in Eqs.~\eqref{eq:na}-\eqref{eq:ne}, respectively, is, however, more tricky to interpret. We notice that if $n_e < n_0/2$ the term changes sign, and we therefore cannot determine the change vector $\textbf{v}$ unambiguously. This hints that there are multiple processes underneath. However, the most obvious separation into $2\gamma_r n_e$ with the $\textbf{v}=(+1,-1)$ and $n_0 n_p$ with $\textbf{v}=(-1,+1)$ permit unphysical transitions. Specifically, if $n_e = n_0$, this choice would allow absorption $n_0 n_p$ to take place, resulting in $n_e = n_0+1$ after the transition - clearly unphysical as we cannot have more excitations than there are emitters present.

The correct decomposition recognizes the contributing physical processes: stimulated emission occurs at rate $\gamma_r n_e n_p$ with $\mathbf{v} = (+1,-1)$, while stimulated absorption occurs at rate $\gamma_r (n_0-n_e)n_p$ with $\mathbf{v} = (-1,+1)$. This ensures that when $n_e = n_0$, the absorption rate vanishes, preserving the constraint $n_e \leq n_0$. The remaining processes in Table~\ref{tab:events} follow directly from the other terms in Eqs.~\eqref{eq:na}-\eqref{eq:ne}.

With the theoretical foundation of the LMC established, together with its relation to the Langevin approach, we now numerically investigate the different approaches over a large parameter regime.

\begin{figure*}
    \centering
    \includegraphics[width=\linewidth]{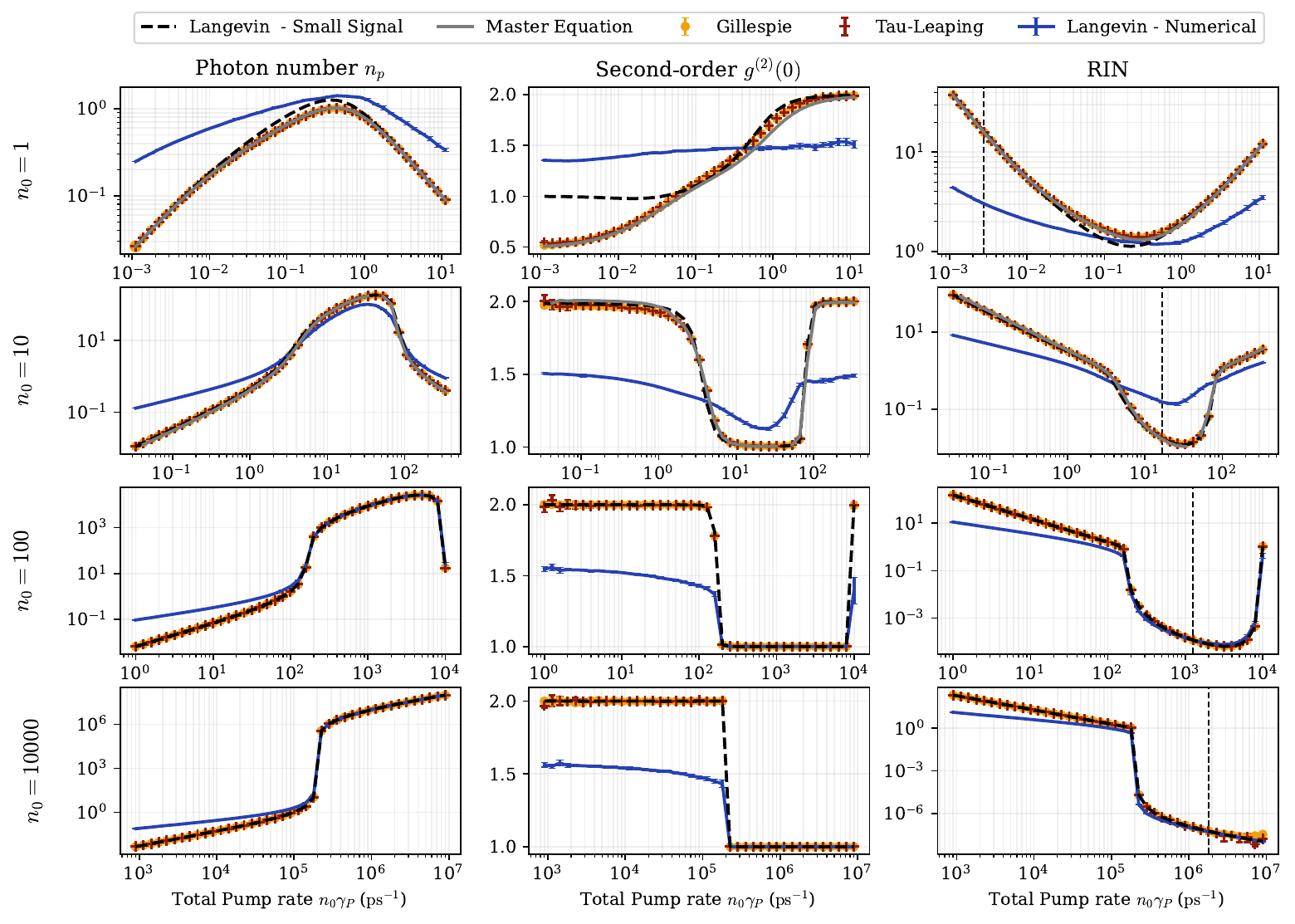}
    \caption{The photon number $n_p$, second-order correlation function $g^{(2)}(0)$, and RIN as function of pump rate for different number of emitters $n_0$ as computed by the analytical small-signal solution to the LRE, the numerical solution to the LRE, the trajectories of the LMC computed using Gillespie First Reaction Method or tau-leaping, and finally for $n_0=1$ and $n_0=10$, the solution obtained using the ME. The parameter values used are light-matter coupling $g=0.1\ \mathrm{ps}^{-1}$, cavity decay rate $\gamma_c = 0.04  \ \mathrm{ps}^{-1}$, pure dephasing rate of the emitters $\gamma_D=1 \ \mathrm{ps}^{-1}$, and a varying non-radiative decay rate for the different emitter numbers $n_0$: $\gamma_A = [0.0, 0.263941, 1.51458, 19.4566] \ \mathrm{ps}^{-1}$. The choice of parameters ensures $\beta = 1/n_0$ as appropriate for investigating the changes from microscopic to macroscopic lasers.}
    \label{fig:comparisons}
\end{figure*}

\section{Numerical comparison \label{sec:numerical}}
In the previous section, we derived the LMC and showed how it converges to the typical laser rate equations with Langevin forces in the many-photon limit. In this section, we compare the numerical solutions to the LMC obtained through Gillespie's method and the tau-leaping method with the numerical solution to the stochastic differential equations constituted by the laser rate equations with stochastic Langevin forces (see appendix \ref{app:numerical_details} for numerical details as well as Ref.~\cite{github} for an open-source implementation). For reference, we also include the analytical small-signal analysis of the Langevin equations, which, in general, apply when the variations in photon and carrier number constitute small relative changes to the steady-state values (see appendix \ref{app:small_signal} for details). 

We consider a varying number of emitters to probe the different regimes of validity of the considered methods. For all simulations, we use the same values for the light-matter coupling rate, $g=0.1\ \mathrm{ps}^{-1}$ \cite{Nomura2010LaserSystem}, the cavity decay rate, $\gamma_c = 0.04  \ \mathrm{ps}^{-1}$ \cite{Dimopoulos2022Electrically-DrivenThreshold}, and the pure dephasing rate of the emitters $\gamma_D=1 \ \mathrm{ps}^{-1}$. However, we vary the non-radiative decay rate of the emitters $\gamma_A$, so that the beta-factor $\beta = \gamma_r(\gamma_P=0)/(\gamma_r(\gamma_P=0) +\gamma_A)$ changes according to $\beta = 1/n_0$. This is a choice we have made so that increasing the number of emitters marks the transition from a microscopic laser with a gradual transition to lasing to a macroscopic laser, where the transition to lasing is sharp; i.e., the photon number changes by orders of magnitude within a small pump interval around the threshold value. With these parameters, the adiabatic elimination and cumulant-expansion used in the derivation of the laser-rate equations and the LMC are good approximations. We note that these are particular choices for the parameter values and are meant to illustrate the transition from smaller to larger lasers and the change in validity of the considered methods. The results highlight important trends but do not establish the absolute regimes of validity of the methods for all parameter values. For example, it has been shown experimentally \cite{Wang2015DynamicalDevices,Wang2020Superthermal-lightLasers} that even for quantum well lasers with beta-factors of $\beta = 10^{-4}$, photon statistics in the transition region exhibit complexity beyond what typical Langevin approaches can capture. However, the LMC correctly predicts the transition region behavior, demonstrating that accurate predictions in the lasing transition region require the LMC framework even for relatively large laser systems.

In Fig.~\ref{fig:comparisons}, we plot the photon number, the same time, second-order auto-correlation function $g^{(2)}(0)$ (see Eq.~\eqref{eq:g2}), and relative intensity noise (RIN) (see Eq.~\eqref{eq:RIN}) as a function of pump-rate $\gamma_P$ and for varying number of emitters $n_0$. For the numerical solution of the Langevin equations, tau-leaping, and Gillespie's method, the plotted values are the average of 5 independent trajectories. For Gillespie's method, the standard deviation of these runs is plotted as error bars. For the numerical Langevin and tau-leaping, we additionally calculate the error due to the timestep $\Delta t$ picked in the numerical solver and include that in the error bars. Many of the error bars are smaller than the line signatures and are therefore barely visible, but nevertheless present. See appendix \ref{app:numerical_details} for details on the trajectory length, choice of time step, and error estimation.

\subsection{Small number of emitters ($n_0=1$)}

In Fig.~\ref{fig:comparisons}, for $n_0 = 1$, we make a number of observations. Due to the small number of emitters, the full ME can be solved numerically. We see that the large dephasing rate $\gamma_D$ makes the system unable to build up a substantial photon population before the radiative rate $\gamma_r$ is quenched due to the pump-induced dephasing \cite{Bundgaard-Nielsen2023StochasticNanolaser,Bundgaard-Nielsen2025SimpleNanolasers}. Thus, the second-order correlation changes from bunched light $g^{(2)}(0)<1$ to thermal light $g^{(2)}(0) \rightarrow 2$, without passing a regime of lasing, where $g^{(2)}(0)=1$ for a range of pump rates. Thus, no lasing takes place.

Gillespie's method and tau-leaping approach are observed to agree well with the simulations of the ME for all pump rates and all statistical measures investigated - the curves are graphically indistinguishable. 

In contrast, both the analytical and numerical solutions to the Langevin equation deviate from the ME. The deviation of the analytical solution is less severe and mostly concerns the $g^{(2)}(0)$-function below threshold and the photon number around lasing. The former is a well-known shortcoming of the Langevin approach \cite{Bundgaard-Nielsen2023StochasticNanolaser}, while the latter is more subtle. As discussed and shown in Appendix \ref{sec:nena}, the deviation originates from the mean-field approximation $\expval{n_e n_p} = \expval{n_p}\expval{n_e}$ made in the LRE, which is generally not the case. This anti-correlation between the emitter and photon number leads to a difference in the mean-photon number, which is perfectly captured by the LMC model. 

For the numerical solution of the Langevin approach, the deviations are more severe, and it is basically invalid for all pump powers, even for the mean photon number. One of the causes for this large deviation between the analytical and numerical solution to the Langevin equations is the possibility of stochastic fluctuations leading to negative emitter or photon populations. To prevent this and to ensure numerical stability, any stochastic fluctuation that tries to reduce a population below zero is instead clamped to zero. The effect of this is a skewing of the noise profile, which causes inaccurate mean populations and statistics. In chemical modeling, this is a known issue and renders the numerical solution to the LRE invalid \cite{Anderson2019OnNetworks}. This problem, of course, only occurs when the populations are very small, but since this is always the case below the lasing threshold, we see that, independently of the emitter numbers $n_0$, the numerical solutions are inaccurate below threshold. Specifically, for $n_0 = 1$, the population is so small for all pump rates that for no pump value, the numerical solution to the LRE is accurate. In appendix \ref{app:numerical_details}, we explore other ways of handling negative populations, but none of these solve the problem.

\subsection{Intermediate number of emitters ($n_0=10$)}

As we increase the emitter number to $n_0 = 10$ in Fig.~\ref{fig:comparisons}, we observe lasing although the threshold transition is not sharp and quenching effects terminate the lasing regime at higher pump rates \cite{Mu1992One-atomLasers,Bundgaard-Nielsen2023StochasticNanolaser,Bundgaard-Nielsen2025SimpleNanolasers}.

Here, for $n_0=10$, it is computationally unfeasible to perform calculations using the standard ME approach.
For identical emitters, one can, however, slightly extend the simulation regime by utilizing the symmetry of the identical emitters. Using the so-called Permutational Invariant Quantum Solver (PIQS) \cite{Shammah2018OpenInvariance} implemented in Qutip \cite{Lambert2024QuTiPPython}, we are able to simulate most of the regime, except when the photon number is too large. As we will see in the following section, the numerical cost of the ME is, however, large: simulation times are more than two orders of magnitude larger than the Gillespie approach.  

Gillespie's method, tau-leaping, and analytical Langevin approaches show excellent agreement with the ME results. We observe only a minor deviation in the analytical Langevin $g^{(2)}(0)$ function near the lasing threshold, likely due to the breakdown of the mean-field approximation $\expval{n_e n_p} \neq \expval{n_p}\expval{n_e}$ \cite{Bundgaard-Nielsen2023StochasticNanolaser} as also discussed in the last section. The numerical solution of the Langevin equations is still invalid for all pump powers for the same reason as above.



\subsection{Large number of emitters ($n_0=100$--$10000$)}

In Fig.~\ref{fig:comparisons}, for $n_0=100$, we are approaching a larger laser system, which is impossible to simulate using a ME, even when exploiting symmetries.
For larger emitter and photon numbers, we thus do not have ME results, and instead establish the validity of our comparisons by two means. First, both previous work and the present study demonstrate excellent agreement between Gillespie's method and ME calculations for systems with up to seven to ten emitters \cite{Bundgaard-Nielsen2025SimpleNanolasers}. This provides confidence in the accuracy of Gillespie’s method within this parameter range, and suggests that it would also be accurate for larger emitter numbers. Second, we focus on the relative performance and systematic differences between methods, which remain meaningful even without absolute ground truth verification. These comparisons reveal regimes where each approach excels or fails, providing useful insights.

With $n_0=100$ and $\beta=0.01$, the transition to lasing is now marked by a sharp kink in the mean photon number as a function of pump rate, as well as the sudden transition of the $g^{(2)}(0)$-function from 2 to 1. We now see that the numerical Langevin approach agrees with the analytical Langevin approach above threshold, which is due to the large photon population obtained above threshold. The numerical solution below the threshold is still invalid due to the low photon number. The tau-leaping, Gillespie's method, and analytical Langevin approach agree very well for all quantities at all pump rates. 

In Fig.~\ref{fig:comparisons}, for $n_0=10000$, the laser is truly macroscopic, and we do not see any signs of quenching. We see the same trend for the numerical approaches, except that Gillespie's approach deviates slightly in the RIN for the largest pump powers. This deviation is due to insufficient statistical sampling. For practical reasons, we limit the Markov-chain simulations to \(8 \times10^{10}\) steps. In the Gillespie approach, large photon numbers lead to very short waiting times between individual events, so that even this number of steps corresponds to a limited simulation time and therefore does not adequately sample the photon statistics. This highlights a different regime where the tau-leaping approach is more efficient than Gillespie's approach, since the tau-leaping approach permits multiple events to occur at each timestep as long as they do not perturb the population significantly, i.e., the populations of emitters and photons are large enough that the changes occurring during the time step do not significantly change the rates appearing in Table \ref{tab:events}. For large photon populations, this means that within a tau-leap, a significant number of events happen, which allows the tau-leaping to sample much more efficiently than Gillespie's approach, which only allows a single event to occur per step. Previously, it was thought that Gillespie's method is always more efficient than the tau-leaping approach \cite{Andre2020}, but we here show that it depends on the parameter regime. We highlight this further in section \ref{sec:time}, where we compare the computational efficiency of the different numerical approaches.


\begin{figure}[!ht]
    \centering
    \includegraphics[width=\linewidth]{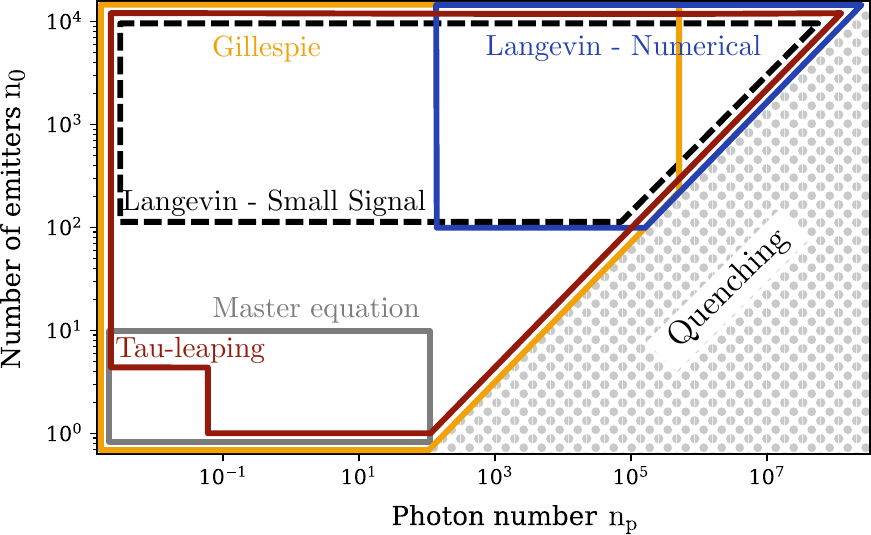}
    \caption{Overview of the validity of methods to compute laser quantum noise as a function of the number of photons and emitters in the laser. The boxes of different colors indicate the areas in which the different methods are valid or computationally feasible. The crossed-out area marks a combination of photon and emitter numbers that are not accessible with the parameters used in this paper due to the physical effect of pump-induced quenching.}
    \label{fig:validity_regimes}
\end{figure}

\subsection{Performance Summary}

In Fig.~\ref{fig:validity_regimes}, we summarize the results of this section by indicating the different regimes of validity of the different methods considered in this section. In a phase diagram of photon and emitter number, we indicate the regimes in which the different methods are either valid or computationally feasible. For example, the ME is only computationally possible for a very small number of emitters and photons. The numerical solution to the LRE, on the other hand, is accurate only for many photons and emitters. Gillespie's method struggles to produce accurate samples for a large number of photons and is thus excluded from this regime. The analytical Small-Signal analysis of the LRE is valid as long as many emitters are present. Finally, the tau-leaping is in principle accurate for all regimes; however, for the smallest number of photons and emitters, it is computationally inefficient, and we have thus excluded it from this microscopic regime. 

Having established where each method produces accurate results, we now examine their computational efficiency, which is a critical consideration for practical applications. In the following section, we compare the time-accuracy tradeoffs to identify which method is optimal for different laser system sizes.

\section{Time VS. accuracy comparisons \label{sec:time}}

\begin{figure}
    \centering
    \includegraphics[width=1\linewidth]{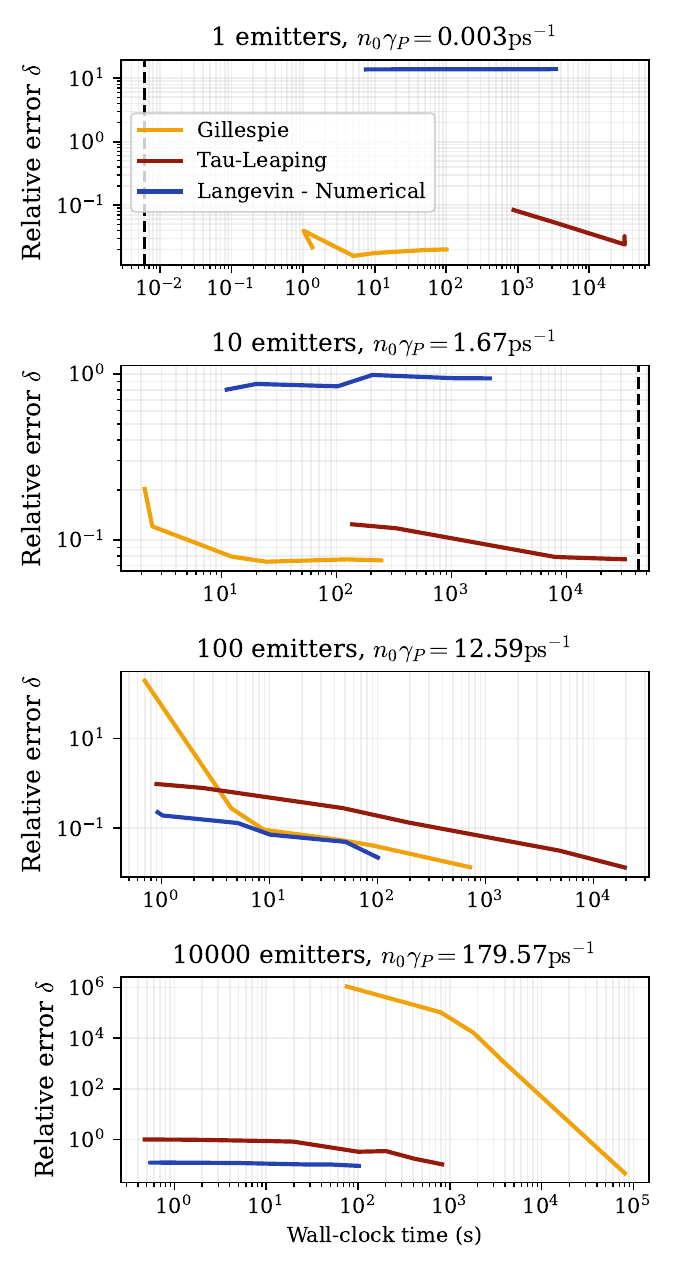}
    \caption{The relative error $\delta$ (largest relative deviation of the three quantities $\expval{n_p}$, $g^{(2)}(0)$, and $\mathrm{RIN}$ over 5 runs) as a function of simulation wall-clock time for different emitter numbers $n_0$ and pump rates $\gamma_P$. The vertical dashed lines indicate the simulation time used for the ME calculations (only for $n_0=1$ and $n_0=10$). For $n_0=1$ and $n_0=10$, the ME results are used as ground truth, while for $n_0=100$ and $n_0=10000$, the analytical Langevin solution is used. Parameters are the same as in Fig.~\ref{fig:comparisons}.
    \label{fig:time_accuracy}}
\end{figure}


In this section, we compare the computational efficiency and accuracy trade-offs of the different numerical approaches. For each emitter number $n_0$ in Fig.~\ref{fig:comparisons}, we select one pump rate (indicated by vertical dashed lines in the RIN plots) and analyze accuracy versus wall-clock time. 

We define the relative error as $\delta^2 = \max( {\expval{\delta_{g^{(2)}(0)}^2}},{\expval{\delta_{RIN}^2}},{\expval{\delta_{n_p}^2}})$, where $\delta_{X} = X_{sim} - X_{truth}$ is the difference between the simulation results $X_{sim}$ (of 5 independent runs) and the ground truth value $X_{truth}$, which for $n_0=1$ and $n_0=10$ is assumed to be the ME results, and for $n_0=100$ and $n_0=10000$ taken to be the analytical small-signal analysis of the LRE. We vary the wall-clock time (the amount of time the simulation has run in real time) by adjusting the number of numerical steps taken in the simulation. The more numerical steps, means more samples and thus the more accurate averages, which leads to a lower relative error. 

In Fig.~\ref{fig:time_accuracy}, we can thus plot the accuracy of the simulation result as a function of wall-clock time for these different emitter numbers $n_0$.


For $n_0=1$, Gillespie's method is both faster and more accurate than the alternatives. The numerical Langevin approach never reaches a high accuracy regardless of the computational time, while tau-leaping requires orders of magnitude more time to achieve similar precision. 

For $n_0=10$, we see similar results, though tau-leaping becomes more competitive with Gillespie in computational efficiency. In the Figure, we also indicate the simulation time of the ME with a vertically dashed line. We see that for $n_0=1$, where the photon population is very small, the ME is much faster than the other methods. For $n_0=10$, there is a more modest photon population, and the numerical cost of the ME is significantly larger than all the other methods. This shows that while it is in principle possible to extend the ME calculation to $n_0=10$, the approach is not scalable beyond this point, and significant performance gains are achieved using the Gillespie or Tau-Leaping approaches

For $n_0=100$, we have picked parameters where we are above threshold with many photons, and thus the numerical Langevin approach is accurate and is also seen to be more efficient with respect to computational time, when compared with the other approaches. As discussed in the previous section, as we increase the population numbers, Gillespie's approach starts to be inefficient as it is simply taking too small steps by only allowing photon and emitter numbers to change by one unit during a time-step. We start to see this decrease in efficiency here, where the Langevin approach is now more efficient. Still, Gillespie's approach is faster than the tau-leaping approach. 
For $n_0=10000$, the tau-leaping approach starts to be more efficient than Gillespie's approach, since it allows multiple events to occur within each step. It is still more computationally intensive when compared with the numerical Langevin approach, however, which motivates the formulation of stochastic differential equations: they are quicker to solve. 

We note that we have picked only particular pump values at each emitter number value. Thus, the conclusion that for $n_0>100$, the LRE outperforms the tau-leaping approach is not generally valid. Specifically, as mentioned previously, if we go below the threshold, the numerical solution to the LRE is invalid, no matter the number of steps, and the appropriate approach would thus here be the tau-leaping or Gillespie's. Still, this section highlights the general trend that for larger laser systems, the LRE becomes more and more efficient when compared with the tau-leaping and Gillespie approach. However, to have an accurate modeling of the complete transition from below to above lasing, it is necessary to use the Gillespie or tau-leaping approaches.

\section{Outlook and Conclusion}

In this paper, we have established the formal connection between the master equation and the so-called stochastic Laser Markov Chain, which treats laser populations as discrete integer variables. We furthermore showed that in the many-photon limit, the Laser Markov Chain simplifies to the Langevin Rate Equations, thereby establishing the validity regimes of different approaches to quantum noise in nanolasers. 

Our numerical comparison of accuracy and computational efficiency revealed that each method has distinct regimes of applicability. Gillespie's method provides the most efficient sampling of the Laser Markov Chain for small populations, while tau-leaping becomes superior for intermediate system sizes. The numerical Langevin approach is only accurate and efficient for very large populations where quantum discreteness effects become negligible.

We also showcase one of the main problems with solving the Langevin Rate Equations numerically; below threshold, when the photon population is small, the Gaussian white noise can cause negative populations. This has to be dealt with by limiting or preventing the solver from producing negative populations, which ultimately skews the noise distribution and causes inaccurate statistics. The analytical solution to the Langevin Rate Equations works for the macroscopic regimes considered in this paper. However, if the laser rate equations are more complicated or large signal modulation needs to be studied, analytical solutions can be difficult, if not impossible, to obtain. The Laser Markov Chain method, on the other hand, has no such restrictions and is numerically stable: it can be used to explore more complex laser systems involving, e.g., non-linear mirrors \cite{Sloan2024Driven-dissipativePhotonics,Nguyen2023IntenseFrequencies,Rivera2023CreatingContinuum,pontula2022strong} or nanolaser modulation, both interesting research directions. Another interesting direction is to include the emitter-photon polarization $\sigma^\dagger a$ directly in the Laser Markov Chain. Langevin equations, which include these correlations, have previously been derived \cite{Maki1989InfluenceSuperfluorescence,Drummond1991QuantumMedium}, and thus it does not seem impossible to derive a Laser Markov Chain which includes such correlations.

\begin{acknowledgments}
This work was supported by the Danish National Research Foundation through NanoPhoton - Center for Nanophotonics, Grant No. DNRF147. The authors are members of the GOLDMINE network.  
\end{acknowledgments}

\appendix

\section{Validity of the Off-diagonal approximation \label{sec:validity}}
When deriving the LMC, we approximated the off-diagonal terms with $\rho_{n_p+1,g;n_p,e} \approx  \left( (n_p-1) / \sqrt{n_p^2+n_p} \right ) \rho_{n_p,g;n_p-1,e}$, based on the recognition that the resulting equations reproduce and mimic the rate equations derived from the ME in Eqs.~\eqref{eq:na}-~\eqref{eq:ne}. We now test the approximation's validity for $n_0 = 1$ -- the only case where we can compare to the explicit result from the ME; all other parameters take the values given in Fig.~\ref{fig:comparisons}.

\begin{figure}[!ht]
    \centering
    \includegraphics[width=\linewidth]{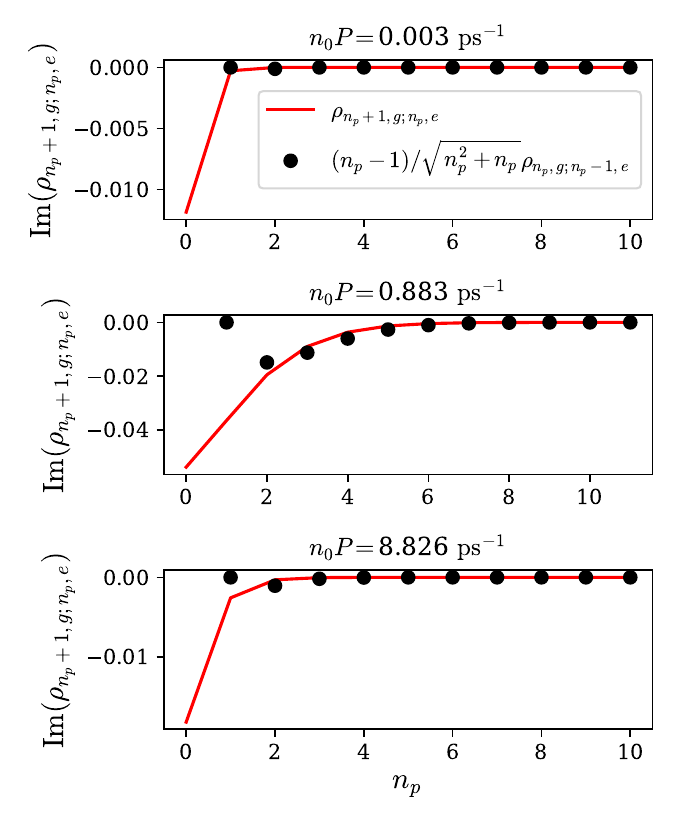}
    \caption{The imaginary part of the off-diagonal element $\rho_{n_p+1,g;n_p,e}$ as calculated using the ME (solid lines), with superimposed dots corresponding to the approximation $\rho_{n_p+1,g;n_p,e} \approx  \left( (n_p-1) / \sqrt{n_p^2+n_p} \right ) \rho_{n_p,g;n_p-1,e}$. For the approximation, the previous matrix element $\rho_{n_p,g;n_p-1,e}$ is obtained from the ME. Parameters are the same as in Fig.~\ref{fig:comparisons} with $n_0 = 1$.}
    \label{fig:offdiagonals}
\end{figure}

In Fig.~\ref{fig:offdiagonals}, we plot the matrix element $\rho_{n_p+1,g;n_p,e}$ as function of $n_p$, together with the approximation $\rho_{n_p+1,g;n_p,e} \approx  \left( (n_p-1) / \sqrt{n_p^2+n_p} \right ) \rho_{n_p,g;n_p-1,e}$. The approximation gives a recursive relation between the elements $\rho_{n_p+1,g;n_p,e}$ and $\rho_{n_p,g;n_p-1,e}$, which we check by using the previously calculated element in the chain of matrix elements $\rho_{n_p,g;n_p-1,e}$ to obtain an expression for $\rho_{n_p+1,g;n_p,e}$.

We choose pump rates which represent the regimes with different $g^{(2)}(0)$, and from the Figure, it is clear that the approximation works well for all pump values for all the matrix elements except for the case $n_p = 1$, where the approximation imposes $\rho_{2,g;1,e} = 0$. Since the only major deviation is for $\rho_{2,g;1,e}$, one expects the approximation to hold well for even larger systems, where the contribution of the first matrix element is smaller. We note that in Fig.~\ref{fig:offdiagonals}, we only plot the imaginary part of the matrix elements, as the real part is zero for all of the matrix elements.

\section{Extending to $n_0$ emitters \label{app:extending}}
In section \ref{sec:LMC}, we derived the main equations for the LMC for only a single emitter. In this section, we extend the result to $n_0$ identical emitters. It is clear that by introducing $n_0$ emitters, we would arrive at a contribution similar to that from Eq.~\eqref{eq:rhong} for each emitter being in the ground state and that from Eq.~\eqref{eq:rhone} from each emitter being excited. We introduce the state $\rho_{n_p,n_e} = \sum \rho_{n_p,x_1,\cdots,x_N}$, with $x_i$ denoting the state of emitter $i$, $n_e$ denoting the number of emitters in the excited state, and the sum runs over all states with $n_e$ excited emitters. We also dropped the double index of the density matrix, and it is now understood that $\rho_{n_p,n_e}$ just refers to a diagonal element. Note that the state $\rho_{n_p,n_e}$ results from a non-trivial combination of $\rho_{n_p,x_1,\cdots,x_N}$ with $n_e$ emitters.

To understand the combinatorics, we start by considering the state with all emitters in the ground state, $\rho_{n_p,0} = \rho_{n_p,g,\cdots,g}$:
\begin{equation}
\begin{aligned}
    &\frac{\partial  \rho_{n_p,0}}{\partial t} = \gamma_r n_p \sum_i^{n_0} \rho_{n_p-1,e_i} - \gamma_r n_p n_0 \rho_{n_p,0} \\
    & + \gamma_c (n_p+1)\rho_{n_p+1,0} - \gamma_c n_p \rho_{n_p,0} + \gamma_A \sum_i^{n_0} \rho_{n_p,e_i} - n_0 \gamma_P \rho_{n_p,0}
\end{aligned}
\end{equation}
where we introduced $\rho_{n_p-1,e_i}$ to represent a single excitation of emitter $i$ (and all other emitters in the ground state). Since, $\rho_{n_p,1} = \sum_i^{n_0} \rho_{n_p-1,e_i}$, we can thus write:
\begin{equation}
\begin{aligned}
    \frac{\partial  \rho_{n_p,0}}{\partial t} &= \gamma_r n_p \rho_{n_p-1,1} - \gamma_r n_p n_0 \rho_{n_p,0} + \gamma_c (n_p+1)\rho_{n_p+1,0} \\
    &- \gamma_c n_p \rho_{n_p,0} + \gamma_A \rho_{n_p-1,1} - n_0 \gamma_P \rho_{n_p,0}
\end{aligned}
\end{equation}
We then want to consider the state $\rho_{n_p,1}$, which consists of all combinations of $\rho_{n_p-1,e_i}$. We start by considering the contribution of one single such state:
\begin{equation}
\begin{aligned}
    \frac{\partial  \rho_{n_p,e_i}}{\partial t} &= \gamma_r n_p \sum_{j \neq i}^{n_0} \rho_{n_p-1,e_i,e_j} - \gamma_r n_p (n_0-1) \rho_{n_p,e_i} \\
    &+ \gamma_c (n_p+1)\rho_{n_p+1,e_1} - \gamma_c n_p \rho_{n_p,e_i} \\
    &+ \gamma_A \sum_{j \neq i}^{n_0} \rho_{n_p,e_i,e_j}  - (n_0-1) \gamma_P \rho_{n_p,e_i} \\ 
    &+ (n_p+1) \gamma_r \rho_{n_p+1,0} - (n_p+1) \gamma_r \rho_{n_p,e_i} \\ 
    &- \gamma_A \rho_{n_p,e_i} + \gamma_P \rho_{n_p,0} \label{eq:rho_e1}
\end{aligned}
\end{equation}
where we now introduced $\rho_{n_p-1,e_i,e_j}$ to denote an excitation in emitter $i$ and $j$. When we combine all contributions into $\rho_{n_p,1}$, most contributions trivially combine to $n_0$ (for example for the term $(n_p+1)\gamma_r \rho_{n_p+1,0}$) or give contributions such as $\sum_i^{n_0} \rho_{n_p,e_i} = \rho_{n_p,1}$ (for example the term $-\gamma_c n_p \rho_{n_p,e_i}$). However, the terms involving multiple excitations, such as $\rho_{n_p-1,e_i,e_j}$, require more careful treatment. We note that, in general, a state consisting of $n_e$ excited emitters with $n_0$ total emitters corresponds to placing $n_e$ indistinguishable balls in $n_0$ containers. The number of combinations, thus, is:
\begin{equation}
    {n_0 \choose n_e} = \frac{n_0!}{n_e!(n_0-n_e)!}, \label{eq:combinatorics}
\end{equation}
Thus, the state $\rho_{n_p,2}$ consists of $n_0(n_0-1)/2$ different terms. When combining Eq.~\eqref{eq:rho_e1} for each possible emitter being excited, we have $n_0$ different equations that each contribute $n_0-1$ terms of the type $\rho_{n_p-1,e_i,e_j}$. These terms will thus combine to form in total $2 \rho_{n_p-1,2}$, and we get:
\begin{equation}
\begin{aligned}
    \frac{\partial  \rho_{n_p,1}}{\partial t} &= 2 \gamma_r n_p \rho_{n_p-1,2} - (n_p+1) \gamma_r \rho_{n_p,1}  \\
    &+ n_0 (n_p+1) \gamma_r \rho_{n_p+1,0} - \gamma_r n_p (n_0-1) \rho_{n_p,1} \\  
    &+ \gamma_c (n_p+1)\rho_{n_p+1,1} - \gamma_c n_p \rho_{n_p,1} +  2 \gamma_A \rho_{n_p,2} \\ 
    &- \gamma_A \rho_{n_p,1} + \gamma_P n_0 \rho_{n_p,0} - (n_0-1) \gamma_P \rho_{n_p,1}     \label{eq:rho_e2}
\end{aligned}    
\end{equation}

To make it general, when we have $n_e$ excited emitters, we will have ${n_0 \choose n_e}$ equation contributions of the type in Eq.~\eqref{eq:rhong} with $(n_0-n_e)$ identical terms that contribute to the state in the higher emitter manifold $\rho_{n_p,n_e+1}$. This gives the factor:
\begin{equation}
\begin{aligned}
    \frac{{n_0 \choose n_e}(n_0-n_e)}{{n_0 \choose n_e + 1}} &= \frac{n_0! (n_e+1)!(n_0-n_e-1)!(n_0-n_e)}{n_0! n_e!(n_0-n_e)!} \\ 
    &= n_e+1
\end{aligned}
\end{equation}
Similarly, we will have ${n_0 \choose n_0-n_e}$ contributions of the type in Eq.~\eqref{eq:rhone}, which each contribute with $n_e$ terms to the state in the lower emitter manifold $\rho_{n_p,n_e-1}$. This gives the factor:
\begin{equation}
\begin{aligned}
   \frac{{n_0 \choose n_0-n_e} n_e }{{n_0 \choose n_e - 1}} &= \frac{n_0! (n_e-1)!(n_0-n_e + 1)! n_e }{n_0! n_e!(n_0-n_e)!} \\ 
   &= n_0 - n_e + 1 = n_g + 1   
\end{aligned}
\end{equation}
Combining all of this, we can write:
\begin{equation}
    \begin{aligned}
        &\frac{\partial  \rho_{n_p,n_e}}{\partial t} = (n_e+1) \gamma_r n_p \rho_{n_p-1,n_e+1} - n_e (n_p+1) \gamma_r \rho_{n_p,n_e} \\
        &+ (n_0-n_e + 1) (n_p+1) \gamma_r \rho_{n_p+1,n_e-1} - \gamma_r n_p (n_0-n_e) \rho_{n_p,n_e} \\  
    &+ \gamma_c (n_p+1)\rho_{n_p+1,n_e} - \gamma_c n_p \rho_{n_p,n_e} \\
    &+  \gamma_A (n_e+1)  \rho_{n_p,n_e+1} - \gamma_A n_e \rho_{n_p,n_e} \\ 
    &+ \gamma_P (n_0-n_e+1) \rho_{n_p,n_e-1} - (n_0-n_e) \gamma_P \rho_{n_p,n_e}. \label{eq:slme_A}
    \end{aligned}
\end{equation}
Thus, we have arrived at Eq.~\eqref{eq:slme} in the main text. 

Finally, we can now use the propensity rates and changes for each process listed in Table \ref{tab:events} and insert them into Eq.~\eqref{eq:cme} to confirm that the LMC samples Eq.~\eqref{eq:slme}.

Considering individually each event type, we obtain, for stimulated emission:
\begin{equation}
    \frac{\partial \rho_{n_p,n_e}}{\partial t} = \gamma_r (n_e+1) (n_p-1)  \rho_{n_p-1,n_e+1}  -  \gamma_r n_e n_p  \rho_{n_p,n_e},\label{eq:cme:st}
\end{equation}
for spontaneous emission, we have:
\begin{equation}
    \frac{\partial \rho_{n_p,n_e}}{\partial t} = \gamma_r (n_e+1) \rho_{n_p-1,n_e+1}  -  \gamma_r n_e \rho_{n_p,n_e}, \label{eq:cme:sp}
\end{equation}
for stimulated absorption, we have:
\begin{equation}
\begin{aligned}
   \frac{\partial \rho_{n_p,n_e}}{\partial t} &= \gamma_r (n_0-n_e+1)(n_p+1) \rho_{n_p+1,n_e-1}  \\
   &-  \gamma_r (n_0-n_e) n_p \rho_{n_p,n_e}, \label{eq:cme:sa} 
\end{aligned}
\end{equation}
for cavity decay, we have:
\begin{equation}
    \frac{\partial \rho_{n_p,n_e}}{\partial t} = \gamma_c(n_p+1) \rho_{n_p+1,n_e}  -  \gamma_c n_p \rho_{n_p,n_e}, \label{eq:cme:cavity}
\end{equation}
for emitter decay, we have:
\begin{equation}
    \frac{\partial \rho_{n_p,n_e}}{\partial t} = \gamma_A(n_e+1) \rho_{n_p,n_e+1}  -  \gamma_A n_e \rho_{n_p,n_e}.
\end{equation}
for pumping, we have:
\begin{equation}
    \frac{\partial \rho_{n_p,n_e}}{\partial t} = \gamma_P(n_0-n_e+1) \rho_{n_p,n_e-1}  -  \gamma_P n_e \rho_{n_p,n_e},\label{eq:cme:pump}
\end{equation}

By adding Eqs.~\eqref{eq:cme:st} and \eqref{eq:cme:sp}, we retrieve the first two terms in Eq.~\eqref{eq:slme_A}. Likewise, Eq.~\eqref{eq:cme:sa} matches the next two terms in Eq.~\eqref{eq:slme_A}, where the identifications continue, chain-like, until all pairs of terms of Eq.~\eqref{eq:slme_A} are matched by Eqs.~\eqref{eq:cme:cavity}-\eqref{eq:cme:pump}. Thus, we have shown a formal connection between the LMC and the ME.

\section{Numerical details \label{app:numerical_details}}
For the simulations of Fig.~\ref{fig:comparisons}, a number of smaller numerical details need examination: how to choose the time step in the tau-leaping and numerical solution to the LRE, how averages are obtained, how errors are estimated, and how negative populations are handled in the numerical solution to the LRE. In this appendix, we address the above. We note that there is also an open-source implementation available to produce all the results of the paper in: \cite{github}.

As mentioned in sec.~\ref{sec:tau_leaping}, the time-step is controlled by the error parameter $\epsilon$ (Eq.~\eqref{eq:tau_leaping_step}) in the tau-leaping algorithm. For $n_0=1,10, \ \mathrm{and} \ 100$ we use $\epsilon = 0.01$, while for $n_0 = 10000$, we use $\epsilon=0.005$.

In the numerical solution of the LRE, we also define an error parameter $\epsilon$, and choose $\Delta t =  \min(\Delta t_a,\Delta t_e)$, where $\Delta t_a = \epsilon^2 n_p / (2 D_{aa})$ and $\Delta t_e = \epsilon^2 n_e / (2 D_{ee})$. The choice of $\epsilon$ thus ensures that the expected relative change in the photon or emitter population due to noise fluctuations is less than $\epsilon$. We use the same numerical values of $\epsilon$ for the numerical solution of the LRE as for the tau-leaping method.

In the numerical solution of the LRE, we choose the total simulation time $T_{Langevin} = 10000/\omega_R$ (with $\omega_R$ defined in Eq.~\eqref{eq:omega_R} to obtain a sufficient sampling of the state space. Instead for the tau-leaping and Gillespie's method, we choose $T_{Markov} = 50 T_{Langevin}$; the extra factor 50 is due to the typically larger stepsize in Gillespie's and tau-leaping methods, which thus requires a longer simulation time to obtain the same number of steps/points for the average. We limit the maximum number of steps to $8 \cdot 10^{10}$ for Gillespie's approach and to $4 \cdot 10^{10}$ for tau-leaping (each step in tau-leaping is more expensive as it requires multiple draws from Poisson distributions). For Gillespie's approach, we also discard the first 10\% of the trajectory for $n_0\geq 100$, as we found the initial state influenced statistics. The initial state was set as the steady state population rounded off to integer values, and this quantization led to a slightly out of equilibrium initial state and a bias in the averaging.

All averages are obtained by:
\begin{equation}
    \expval{ X } = \frac{1}{T_{tot}} \sum_i^N X(t_i) \Delta t_i,
\end{equation}
where $T_{tot}$ is the total simulation time, $X(t_i)$ is the variable at time $t_i$, $\Delta t_i$ is the timestep size at time $t_i$. For LRE and tau-leaping, $t_i$ is the same for all time steps, while for Gillespie's method, it varies.

\begin{figure*}[!ht]
    \centering
    \includegraphics[width=\linewidth]{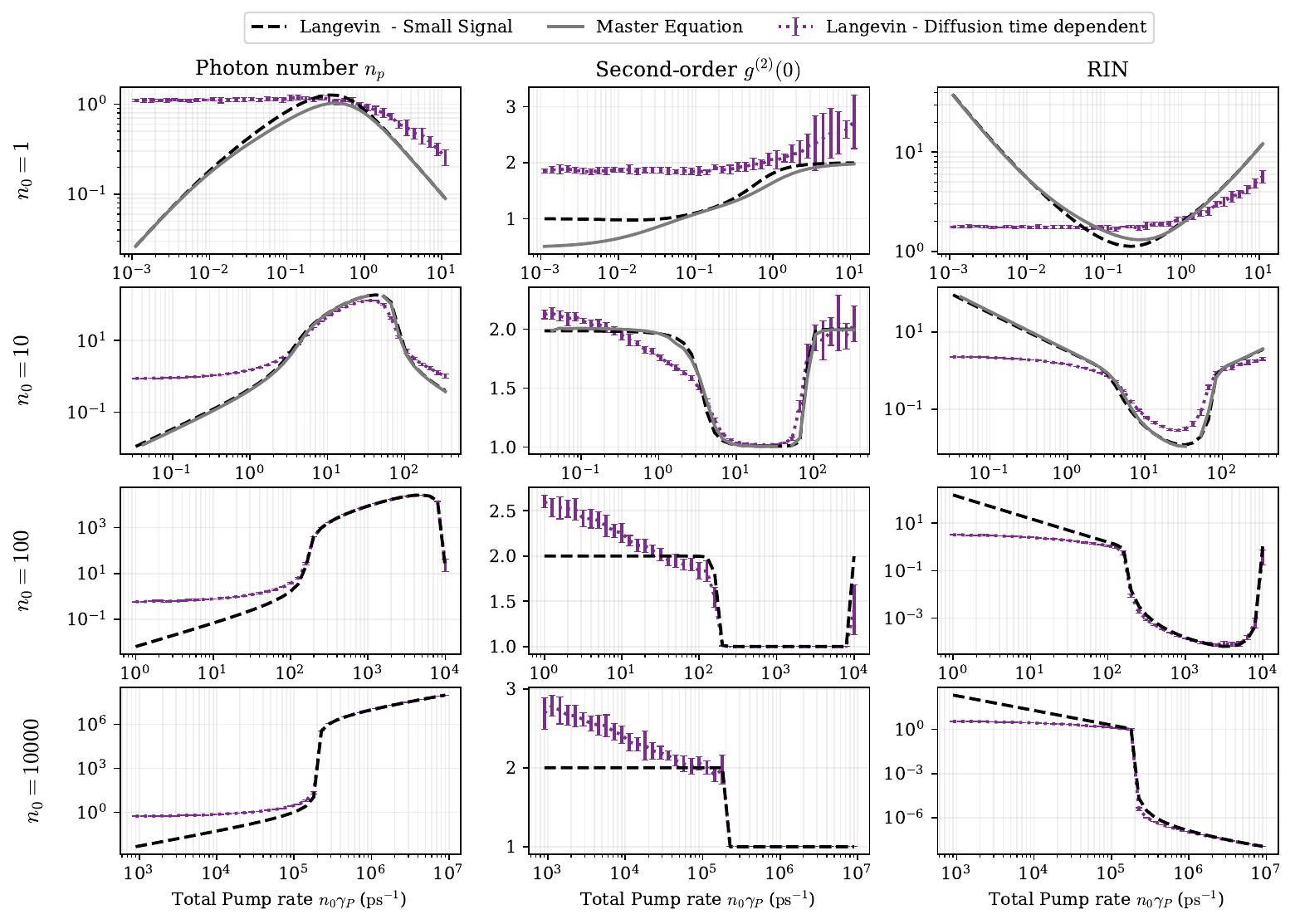}
    \caption{Same as Fig.~\ref{fig:comparisons}, but showcasing different numerical solutions to the Langevin rate equations. Langevin - Diffusion time dependent (purple) denotes the numerical solution where the diffusion terms of the stochastic differential equation are updated at each timestep as opposed to just using the average diffusion. }
    \label{fig:comparison_langevin}
\end{figure*}

To estimate the error in the simulations, we need to take into account the error due to stochastic variations and, in the tau-leaping and numerical solution to LRE, the error due to the finite time step. To estimate stochastic errors, we run 5 trajectories and take the standard deviation of the resulting statistical averages (photon number $n_p$, $g^{(2)}(0)$, and RIN). To estimate the error due to the timesteps we run 5 trajectories with a timestep twice the previous one and include the error $\delta_{\Delta t}$ in the error bars as $\delta = \sqrt{\delta_{\Delta t}^2 + \mathrm{STD}}$: $\mathrm{STD}$ is the standard deviation of the 5 runs with the lowest time step $\Delta t$, $\delta_{\Delta t} = \abs{\Bar{Y}_{\Delta t = 2}-\Bar{Y}_{\Delta t = 1}}$ is the upper limit on the error induced by the larger timestep, where $\Bar{Y}_{\Delta t = T}$ is the expectation value of the quantity $Y$ as found from the 5 runs with the time step $\Delta t = T \delta$ ($T=1$ means thus means a smaller timestep than $T=2$). These are the error bars shown in Fig.~\ref{fig:comparisons} and \ref{fig:comparison_langevin}.

As mentioned in the main text, the stochastic noise in the LRE can lead to negative populations. The negative populations lead to ill-defined rates and numerical instabilities. To combat this, we limit each step in the differential equation solver with a "discrete callback" which ensures that the population is positive. The results shown in the main text clamp the variables so that $n_p = \max(0,n_p)$ and $n_e = \max(0,n_e)$ to avoid any negative populations. We also ensure $n_e \leq n_0$ by clamping $n_e = \min(n_0,n_e)$. We have also investigated an approach where instead of clamping the populations, we use $n_p = \abs{n_p}$ and $n_e = \abs{n_e}$, corresponding to "reflections", if any population is negative. This does not impact the result meaningfully, as the noise profile is still skewed, leading to wrong predictions.

In Fig.~\ref{fig:comparison_langevin}, we also include numerical results using a different computational implementation of the stochastic diffusions in Eq.~\eqref{eq:sde} and Eqs.~\eqref{eq:d_aa}-\eqref{eq:d_ee}. So far, in all the results shown, the rates of the diffusion terms were based on the steady-state populations $\Bar{n}_a$ and $\Bar{n}_e$. This means that the diffusion terms of Eqs.~\eqref{eq:d_aa}-\eqref{eq:d_ee} were evaluated once at the beginning of the simulation by finding the steady state solution to the (noiseless) rate equations. Again, this was for the purpose of numerical stability. In Fig.~\ref{fig:comparison_langevin}, we, however, also show the results using diffusion terms that are updated at each timestep, i.e., not based on the average populations. That means the noise is now multiplicative and the stochastic differential equation is instead solved using an Euler-Heun scheme implemented in DifferentialEquations.jl in Julia \cite{christopher_rackauckas_2025_17427545}, whereas before an Euler-Maruyama scheme from the same library was used. We still employ the clamping of populations, but from Fig.~\ref{fig:comparison_langevin}, it is clear that the numerical instabilities lead to diverging behaviors and still substantial errors.

If external temporal modulations of the pumping rate were to be considered, the diffusion coefficients would have to be updated at each time step to reflect the changing pump rate and thus average populations, but the results show that this only aggravates the instability. 

\section{Small-Signal Analysis \label{app:small_signal}}
We consider the stochastic differential equations arising due to the addition of Langevin forces in the rate equations:
\begin{align}
      &\frac{d n_p}{dt} = \gamma_r (2n_e-n_0)n_p + \gamma_r n_e- \gamma_c n_p + F_{n_p}, \label{eq:f_na} \\
      &\frac{d n_e}{dt} = \gamma_P (n_0-n_e) - \gamma_r (2n_e-n_0)n_p -\gamma_r n_e - \gamma_A n_e+ F_{n_e}. \label{eq:f_ne}
\end{align}
The Langevin forces $F_{n_p}$ and $F_{n_e}$ follow the definitions in Eqs.~\eqref{eq:f_whitenoise}-\eqref{eq:d_ee}. Here, the small-signal analysis offers the advantage of analytical expressions that circumvent the inaccuracies of numerical integration at low photon numbers (section \ref{sec:numerical}).

It is often customary to analyze the Langevin rate equation using small-signal analysis to achieve analytical expressions instead of solving the stochastic differential equations numerically. As we saw in section \ref{sec:numerical}, the numerical solutions become inaccurate for small photon numbers since noise fluctuations cause negative populations. Although free from the numerical constraints, the analytical solution, however, can only describe small perturbations, and thus, large pump modulations cannot be treated this way. The starting point of the analysis is the linearization of Eqs.~\eqref{eq:f_na}-\eqref{eq:f_ne} around the steady state values $\Bar{n}_a$ and $\Bar{n}_e$ \cite{Coldren1997}:
\begin{equation}
\frac{d}{dt}
\begin{pmatrix}
\Delta n_p \\
\Delta n_e
\end{pmatrix}
=
\begin{pmatrix}
-\Gamma_{aa} &
\;\Gamma_{ae} \\
-\Gamma_{ea} &
\;-\Gamma_{ee}
\end{pmatrix}
\begin{pmatrix}
 \Delta n_p \\
\Delta n_e
\end{pmatrix}
+
\begin{pmatrix}
F_{n_p} \\
F_{n_e}
\end{pmatrix},
\label{eq:linear_langevin}
\end{equation}
where
\begin{align}
    & \Gamma_{aa} = \gamma_c - \gamma_r(2\Bar{n}_e-n_0), \\
    & \Gamma_{ae}  = 2\gamma_r \Bar{n}_{a}+\gamma_,r\\
    & \Gamma_{ea} = \gamma_r(2n_{e0}-n_0), \\
    & \Gamma_{ee} = \gamma_P + 2\gamma_r n_{a0} + \gamma_r+ \gamma_A.
\end{align}
Fourier transforming Eq.~\eqref{eq:linear_langevin}, we obtain:
\begin{align}
    &\Delta n_p(\omega) = \frac{H(\omega)}{\omega_R^2}( F_{n_e}(\omega)(\Gamma_{aa} + i \omega)- F_{n_p}(\omega) \Gamma_{ea}), \\
    &\Delta n_e(\omega) = \frac{H(\omega)}{\omega_R^2}( F_{n_p}(\omega)(\Gamma_{ee} + i \omega) + F_{n_e}(\omega) \Gamma_{ae}),
\end{align}
with $H(\omega)= \omega_R^2/(\omega_R^2 - \omega^2 + i \omega \Gamma)$ and with
\begin{equation}
    \omega_R = \Gamma_{ae}\Gamma_{ea} + \Gamma_{aa}\Gamma_{ee}, \label{eq:omega_R}
\end{equation}
and $\Gamma = \Gamma_{aa} + \Gamma_{ee}$. The Fourier transformed Langevin forces obey:
\begin{equation}
    \expval{F_i(\omega)F_j(\omega')} =  2 D_{ij} \delta(\omega-\omega'), 
\end{equation}
With this, we get:
\begin{align}
    &S_{n_p}(\omega) = \frac{1}{2\pi} \int_{-\infty}^\infty d \omega' \ n_p(\omega)n_p(\omega') \\
    & = \frac{|H(\omega)|^2}{\omega_R^4}\left ( 2 \Gamma_{ae}^2 D_{ee} + 4 \Gamma_{ee}\Gamma_{ae} D_{ae} + (\Gamma_{ee} + \omega^2)D_{aa} \right),
\end{align}
from which we can compute the photon variance:
\begin{equation}
\begin{aligned}
    &\expval{\Delta n_p^2}  = \frac{1}{2\pi} \int_{-\infty}^\infty d \omega \ S_{n_p}(\omega) \\
    & \frac{1}{\Gamma}\left[\left(1+\frac{\Gamma_{e e}^2}{\omega_R^2}\right) D_{aa}+\frac{\Gamma_{a e}^2}{\omega_R^2} D_{e e}+\frac{2 \Gamma_{a e} \Gamma_{e e}}{\omega_R^2} D_{a e}\right],
\end{aligned}    
\end{equation}
which we can use to subsequently compute the RIN and $g^{(2)}(0)$ using:
\begin{align}
    & g^{(2)}(0) =  \frac{\expval{n_p^2} - \expval{n_p}}{\expval{n_p}^2} \label{eq:g2} \\
    &  \mathrm{RIN} =  g^{(2)}(0) +\frac{1-\expval{n_p}}{\expval{n_p}}  \label{eq:RIN}
\end{align}
where $\expval{n_p^2} = \expval{n_p}^2 + \expval{\Delta n_p^2}$.

\section{Validity of meanfield approximation \label{sec:nena}}
\begin{figure}[H]
    \centering
    \includegraphics[width=\linewidth]{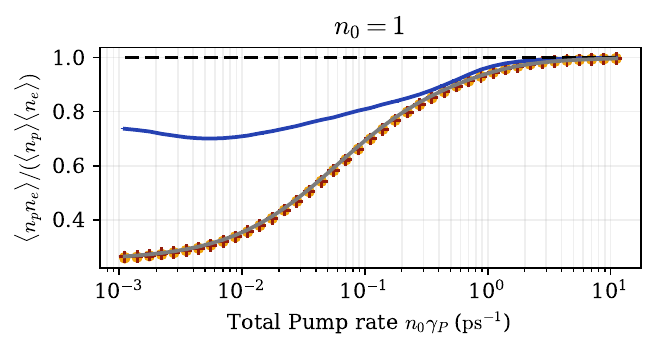}
    \caption{The fraction of deviation from the mean field approximation $\expval{n_p n_e}/(\expval{n_e}\expval{n_p})$ as function of pump rate for $n_0=1$ as predicted by the various methods. Other parameters and legends are the same as in Fig.~\ref{fig:comparisons}.}
    \label{fig:nena}
\end{figure}

As we saw in Fig.~\ref{fig:comparisons}, the steady state solution to the rate equations produced deviations in the photon number for $n_0=1$, when compared with the ME. For all other emitter numbers, the rate equations match the prediction of Gillespie and tau-leaping methods. We can understand this deviation by considering the emitter-photon correlation $\expval{n_p n_e}$, which in the steady-state solution is assumed to be $\expval{n_p n_e} = \expval{n_p}\expval{n_e}$, i.e., a mean-field approximation. In Fig.~\ref{fig:nena}, we plot $\expval{n_p n_e}/(\expval{n_e}\expval{n_p})$ as a function of pump rate for $n_0=1$ for the different methods considered in Fig.~\ref{fig:comparisons}. A deviation from unity, thus, means that the mean field approximation is violated. We see that the ME, Gillespie, and tau-leaping all agree and predict a significant deviation from the mean-field assumption for almost all pump rates. As quenching is reached for large pump rates, the mean field approximation is, however, approached. This is in contrast to the steady-state solution, which assumes $\expval{n_p n_e}/(\expval{n_e}\expval{n_p}) = 1$ for all pump rates, and this explains the deviation seen in Fig.~\ref{fig:comparisons}.

\end{document}